\documentclass[twocolumn,showpacs,preprintnumbers,amsmath,amssymb]{revtex4}
\usepackage{graphicx}
\usepackage{dcolumn}
\usepackage{bm}
\usepackage{amsmath}
\usepackage{amsfonts}%

\begin{document}

\title{Theoretical study of radiative electron attachment to CN, C$_2$H, and C$_4$H radicals}
\author{Nicolas Douguet$^{1}$, S. Fonseca dos Santos$^{1}$, Maurice Raoult$^{2}$, Olivier Dulieu$^{2}$, Ann E. Orel$^{1}$, and Viatcheslav Kokoouline$^{3}$}
\affiliation{$^{1}$ Department of Chemical Engineering and Materials Science, University of California at Davis, Davis, CA 95616, USA \\$^{2}$Laboratoire Aim\'e Cotton, CNRS/Universit\'e Paris-Sud/ENS Cachan, b\^at. 505, Campus d'Orsay, 91405 Orsay cedex, France\\$^{3}$Department of Physics, University of Central Florida, Orlando, Florida 32816, USA }


\begin{abstract} 
A first-principle theoretical approach to study the process of radiative electron attachment is developed and applied to the negative molecular ions CN$^-$, C$_4$H$^-$, and C$_2$H$^-$. Among these anions, the first two have already been observed in the interstellar space. Cross sections and rate coefficients for formation of these ions by radiative electron attachment to the corresponding neutral radicals are calculated. For completeness of the theoretical approach, two pathways for the process have been considered: (i) A direct pathway, in which the electron in collision with the molecule spontaneously emits a photon and forms a negative ion in one of the lowest vibrational levels, and (ii) an indirect, or two-step pathway, in which the electron is initially captured through  non-Born-Oppenheimer coupling into a vibrationally resonant excited state of the anion, which then stabilizes by radiative decay. We develop a general model to describe the second pathway and show that its contribution to the formation of cosmic anions is small in comparison to the direct mechanism. The obtained rate coefficients at 30~K are $7\times 10^{-16}$cm$^3$/s for CN$^-$, $7\times 10^{-17}$cm$^3$/s for C$_2$H$^-$, and $2\times 10^{-16}$cm$^3$/s for C$_4$H$^-$. These rates weakly depend on temperature between 10K and 100 K. The validity of our calculations is verified by comparing the present theoretical results with data from recent photodetachment experiments. 
\end{abstract}


\maketitle

\section{Introduction}

The present theoretical study of radiative electron attachment (REA) to neutral molecules is motivated by the recent discoveries of molecular anions in the interstellar medium (ISM). Six anions have been detected so far in the ISM: C$_6$H$^-$  \cite{mccarthy06,cernicharo07,herbst08,harada08,cernicharo08}, C$_4$H$^-$ \cite{gupta07}, C$_8$H$^-$ \cite{kawaguchi07}, C$_3$N$^-$ \cite{thaddeus08}, C$_5$N$^-$ \cite{cernicharo07,cernicharo08}, and CN$^-$ \cite{agundez10}. The possibility for atomic anions, such as H$^-$, to be formed in the ISM by REA was first suggested by McDowell \cite{mcdowell61} in 1961. Later, Dalgarno and McCray  \cite{dalgarno73} have discussed the role of negative atomic ions in the formation of neutral molecules in the ISM. The formation of  molecular anions in the ISM by REA has been proposed by Herbst \cite{herbst81}, who has also developed a theoretical approach \cite{herbst81,herbst08} to evaluate rate coefficients for REA. More than twenty years after his prediction, negative molecular ions were indeed detected in the ISM. 

The theoretical approach proposed by Herbst, the phase-space theory (PST), has been used in a number of studies \cite{herbst81,herbst08,herbst85,terzieva00,petrie97,millar07,petrie96}  to calculate the REA rate coefficients and to model formation of anions in the ISM. The approach has been successful in interpreting the observed column density of  C$_8$H$^-$, C$_6$H$^-$, C$_5$N$^-$ and C$_3$N$^-$ ions, while the agreement with observations is not as good for the C$_4$H$^-$ and CN$^-$ ions. 

\begin{figure}
\includegraphics[width=8cm]{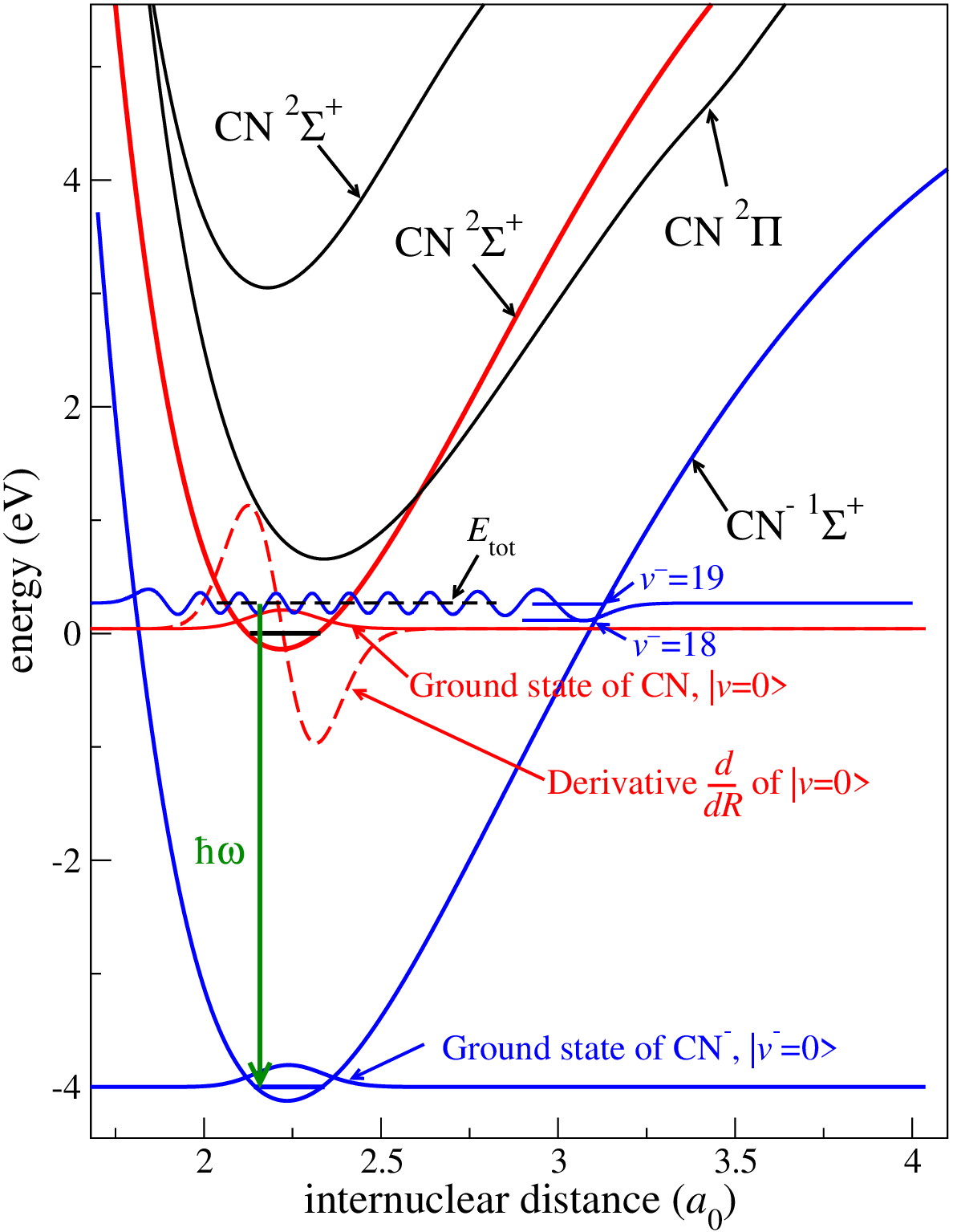}
\caption{Schematic representation of two mechanisms for REA of an electron to the CN radical: (i) DREA: The electron incident on CN in its ground vibronic state spontaneously emits a photon of energy $\hbar\omega$ (green arrow) and forms the CN$^-$ ground state. The photon energy is equal to the difference between the initial total energy of the system $E_{tot}$ (horizontal dashed line at near energy of the $v^-=19$ state) and the energy of CN$^-$ ground state. (ii) IREA: As a first step of the process, the incident electron is captured via non-Born-Oppenheimer coupling into the ground electronic state of CN$^-$ without emitting a photon. Because the total energy $E_{tot}$ is conserved, the vibrational level $v^-$ of CN$^-$ in this first step of the process is highly excited: For low incident energies of the electron, it corresponds to $v^-=18$ or 19. After the first step, the electron captured in an excited vibrational level can either autodetach or stabilize by photon emission. The photon changes the rotational and vibrational  states of the CN$^-$ molecule. The largest probability for the second step corresponds to a change of $v^-$ by one quantum. The dashed red line shows the derivative of the vibrational ground state wave function of CN, which enters in the calculation of the non-Born-Oppenheimer coupling in Eq. (\ref{eq:nac1}).} 
\label{fig:RA_indirect}
\end{figure}

PST relies on several assumptions. First, it considers  REA as a two-step process, schematically represented in Fig. \ref{fig:RA_indirect} for the case of CN/CN$^-$. As a first step, the incident electron is captured by a neutral target molecule $M$ into an electronic state of $M^-$ through non-Born-Oppenheimer coupling, thus forming a vibrational resonance of $M^-$ in a highly-excited vibrational level. The resonance can decay back to the $M+e^-$ electronic continuum spectrum through the same non-Born-Oppenheimer coupling, i.e. the electron can autodetach. Another possibility for the resonance, considered as the second PST step, is to emit a photon, thus stabilizing the $M^-$ system. In PST, it is assumed that the probability of the first step of the process is unity and, therefore, the cross section for the electron capture is approximated by the unitary limit  formula for  $s$-wave  scattering  \cite{herbst08} $\sigma_c=\pi/k^2$, 
where $k$ is the wave number of the incident electron with energy $E_{el}=(\hbar k)^2/(2m_e)$, and $m_e$ is the electron mass. In the second PST step, stabilization of the $M^-$ resonance  is represented as an emission of a photon by a set of harmonic oscillators of the molecule in the normal mode approximation of molecular vibrations. A larger number of available vibrational modes decreases the probability of autodetachment, and thus allows for a larger probability in the second PST step. Therefore, in PST, the overall two-step REA cross section $\sigma_{PST}$ grows rapidly with the number of atoms in the molecule and approaches the unitary limit $\sigma_{PST}\to \sigma_c$. For example, the REA rate coefficient for formation of CN$^-$ calculated by PST is about $10^{-17}$cm$^3$/s  \cite{petrie96} and much larger, $3\times 10^{-7}$cm$^3$/s, for C$_6$H$^-$ \cite{herbst08} at 10K. For CN$^-$ the theoretical value is smaller by several orders of magnitude than the one needed to explain the [CN$^-$]/[CN] abundance ratio obtained from the astrophysical observations. 

The PST assumption about the unitary probability of the first step of the process can hardly be justified for small molecular ions. But it is possible to apply a quantum-mechanical approach based on first principles for the both steps of the REA mechanism, suggested by Herbst. Considering the process quantum-mechanically, another mechanism for REA is also possible: The electron incident on a neutral molecule emits a photon and becomes bound without an intermediate step of capture into a vibrationally excited state of the ground electronic state of the negative ion. To distinguish the two mechanisms, we call the first one as indirect (IREA) and the second mechanism as direct REA (DREA). The total cross section of the REA is the sum of DREA and IREA cross sections.

Recently, we have developed a fully-quantum theoretical approach for DREA based on first principles only \cite{douguet13}. The approach considers the radiative electron attachment  of a continuum electron through spontaneous decay to the anion ground state (see Fig. \ref{fig:RA_indirect}) and does not include an intermediate vibronic state of $M^-$ populated through the non-Born-Oppenheimer coupling, as represented by the two-step REA mechanism in PST.  Hence, the approach relies on wave functions of the continuum spectrum of the $M+e^-$ system and transition dipole moments to the bound electronic state of $M^-$, calculated {\it ab initio}. Using the developed approach, we have calculated the REA cross section and rate coefficient for the formation of the cyanide ion, CN$^-$. Our results confirmed the previous assessment that the REA rate coefficient for CN$^-$ is too small,  $8\times 10^{-16}$cm$^3$/s at 10~K \cite{douguet13}, to explain the CN$^-$ abundance observed in the ISM. 

In the present study, we extend the theoretical approach to  larger molecules.  In order to assess the approximations employed in the PST approach, we also develop a quantum mechanical approach for IREA.  In this study, we calculate explicitly the probability of electron capture during the first step of IREA process using {\it ab initio} methods and determine the overall cross section of the IREA process, which might be compared with the PST results.

Although there is no experimental data on the REA process for carbon chain molecules, using a similar theoretical approach, we determine in this study cross sections for the inverse process to REA, namely photodetachment (PD), and compare with available data  from recent photodetachment experiments \cite{kumar13,best11}. This  allows us to verify the validity of our results.

In the next section, we present our theoretical approach to study DREA and apply it to the CN$^-$, C$_2$H$^-$, and  C$_4$H$^-$ ions. Section \ref{sec:photodetachment} is devoted to the comparison of the results obtained in this study with data from photodetachment experiments. Section \ref{sec:IREA-CN} presents the theoretical approach of IREA for the case of CN$^-$. In the Section \ref{sec:IREA-larger}, we develop a model of IREA for larger molecules and we compare its results with those of PST. Finally, Section \ref{sec:conclusion} summarizes the important findings of the study.

\section{Direct mechanism of radiative electron attachment to CN, C$_2$H, and C$_4$H}
\label{sec:theory}


\begin{figure}
\includegraphics[width=7cm]{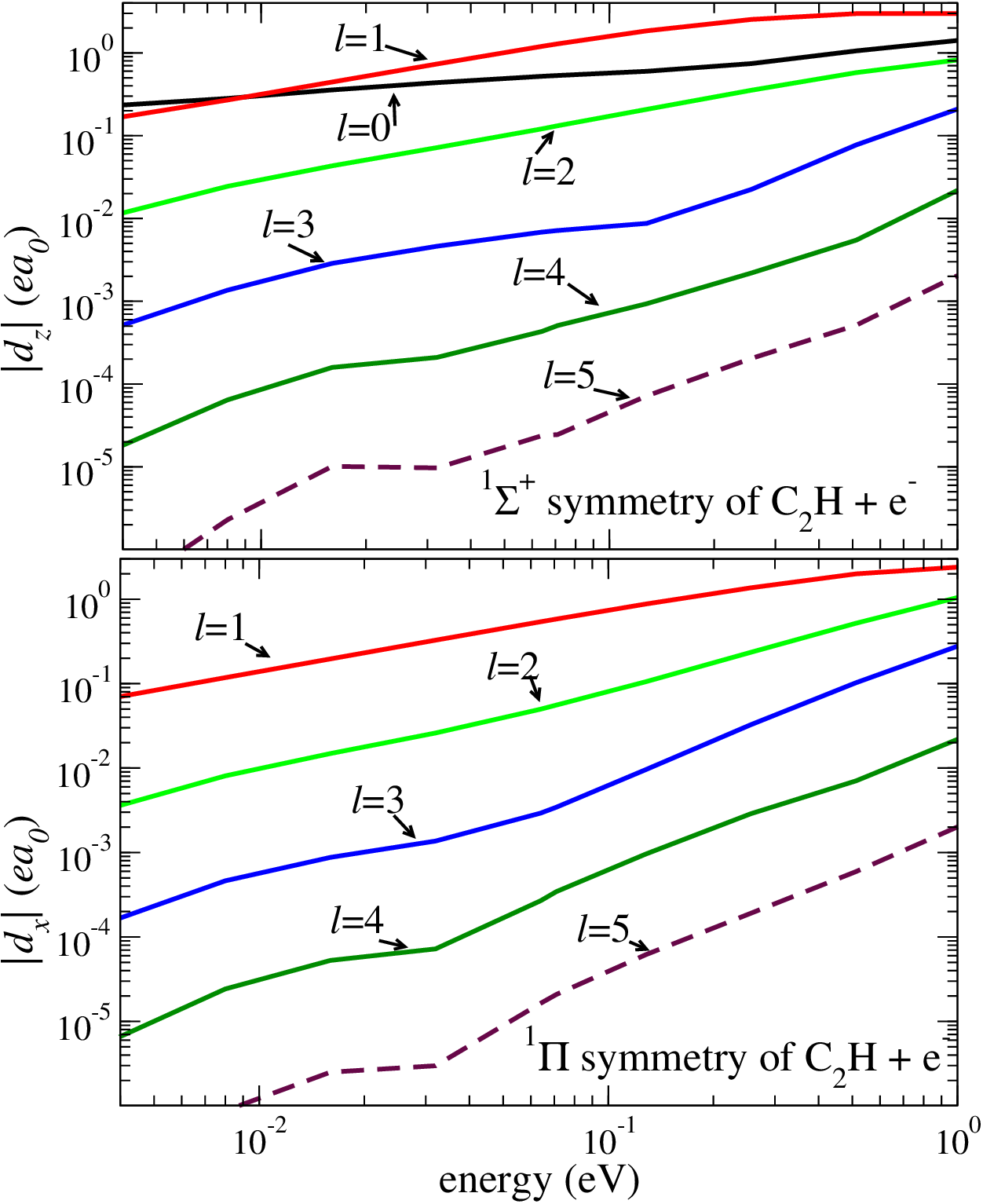}
\caption{Matrix elements of the $ d_x$ and $d_z$ components of the transition dipole moment between the C$_2$H$^-$ electronic ground state and the $e^-+$C$_2$H system for several partial waves as a function of the incident electron energy, calculated at the equilibrium geometry of C$_2$H$^-$.}
\label{fig:D_E_C2H}
\end{figure}

The cross section for DREA of an electron to a neutral linear molecule $M$, such as C$_n$H ($n=2,4$) or C$_m$N ($m=1,3$), initially in its electronic ground state with the vibrational level $v$ and energy $E_i$, was given in \cite{douguet13} and is expressed as
\begin{equation}
\label{eq:cs_RAE2}
\sigma_{i}=\frac{4}{3}\frac{\pi \omega^3 m_e}{k^2\hbar^2 c^3} \sum_{l\pi} \left| d_{\pi,\Gamma l-\pi}^{(v\to v_f)}\right|^2\,,
\end{equation}
where $v_f$ is the vibrational state of the ion $M^-$ with total energy $E_f$ formed after DREA; $\omega$ is the frequency of the emitted photon, $\hbar\omega=E_i+E_{el}-E_f$. The quantities $d_{\pi,\Gamma l\lambda}^{(v\to v_f)}$ are matrix elements of the components $\pi=-1,0,+1$ of the dipole moment operator  between the initial $\Psi_{\Gamma l\lambda} $ ($M$+incident electron) and the final $\Psi_f $ electronic state, integrated over the initial $\chi_v(\vec{q})$ and final $\chi_{v_f}(\vec{q})$ vibrational wave functions of $M$ and $M^-$, respectively, and $\vec{q}$ denotes collectively all internuclear degrees of freedom. Their values are given by
\begin{eqnarray}
\label{eq:vib_tdm}
d_{\pi,\Gamma l\lambda}^{(v\to v_f)}=\int \chi_{v_f}(\vec{q})\langle\Psi_f\vert d_\pi\vert \Psi_{\Gamma l\lambda}\rangle \chi_{v}(\vec{q})d\vec{q} \,,
\end{eqnarray}
(see details in Ref. \cite{douguet13}); with $l$ the electronic partial wave angular momentum and $\lambda$ its projection on a specific axis in the molecular frame. The matrix elements $\langle  \Psi_f\arrowvert d_\pi\arrowvert \Psi_{\Gamma l\lambda}\rangle$ of the dipole moment operator $\hat d_\pi$ are given by the integral
\begin{eqnarray}
\label{eq:transition-dipole}
\langle  \Psi_f\arrowvert d_\pi\arrowvert \Psi_{\Gamma l\lambda}\rangle&=&-\sum^{N}_{k=1}\int\Psi_f^*(r_1,\cdots, r_N)er_{k\pi} \nonumber\\
&&\times\Psi_{\Gamma l\lambda}(r_1,\cdots, r_N)
 d^3r_1\cdots d^3r_N,
\end{eqnarray}
where the function $\Psi_f$ represents the $N$-electron final state of the negative ion ($r_1$,...,$r_N$ are the coordinates of the electrons), $\Psi_{\Gamma l\lambda}$ is the electronic continuum state representing the scattering electron with $l$ and $\lambda$ angular quantum numbers, and $\Gamma$ labels the initial neutral electronic target state with $N-1$ electrons. Finally, $r_{k\pi }$ is one of the three cyclic components ($\pi=0,\pm 1$) of the coordinate of the $k^\mathrm{th}$ electron
\begin{equation}
 r_{k\pi }= \left\{\begin{array}{l}
z_k,~{\rm if}~\pi=0\\
\mp(x_k\pm iy_k)/\sqrt{2},~{\rm if}~ \pi=\pm 1 \,.
\end{array}
\right. 
\end{equation}
The calculations of the electronic wave functions and transition dipole moments (TDMs) of Eq. (\ref{eq:transition-dipole}) are performed using the complex Kohn variational method, extensively described in past studies \cite{mccurdy89,orel91}. 


The DREA cross section for the formation of CN$^-$, starting from the ground vibrational level $v=0$ of CN, was calculated using Eq. (\ref{eq:cs_RAE2}) \cite{douguet13}. The vibrational integral of Eq. (\ref{eq:vib_tdm}) was computed explicitly from the geometry-dependent matrix elements $\langle\Psi_f\vert d_\pi\vert \Psi_{\Gamma l\lambda}\rangle$ obtained in the complex Kohn calculations. These matrix elements, as a function of the electron energy and of the internuclear distance, were respectively shown in Fig. 3 and Fig. 4 of Ref. \cite{douguet13}.

\begin{figure}[b]
\includegraphics[width=7cm]{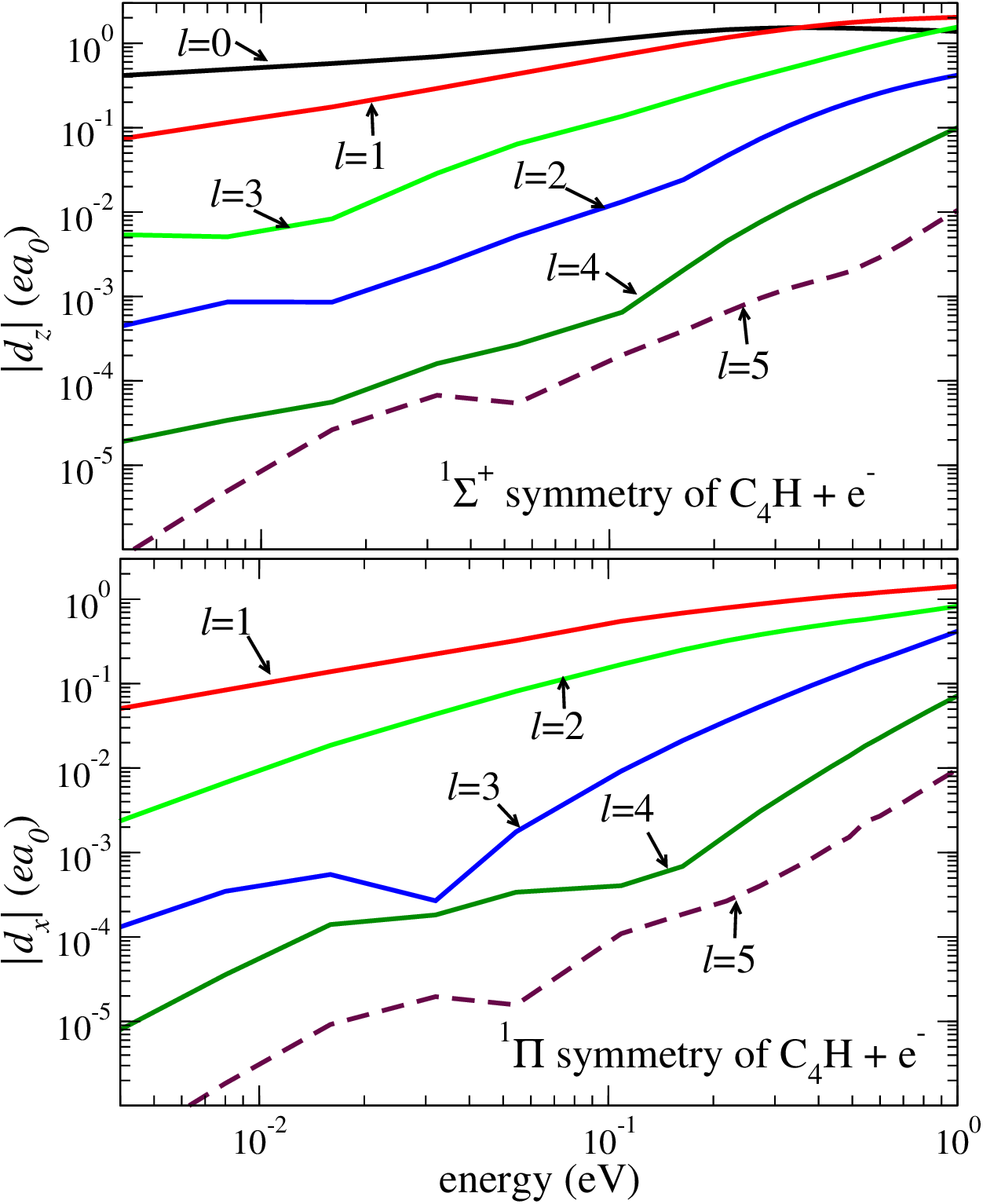}
\caption{Same as Fig. \ref{fig:D_E_C2H} for the  $e^-+$C$_4$H system.}
\label{fig:D_E_C4H}
\end{figure}

The calculation of the TDMs is computationally intensive, especially for polyatomic molecules. In Ref. \cite{douguet13}, we found that the TDMs weakly depend on the geometry of the molecule near its equilibrium position. Therefore, it seems reasonable to use the Franck-Condon approximation and simplify the calculation of the vibrational integral in Eq. (\ref{eq:vib_tdm}) by using the value of $\langle\Psi_f\vert d_\pi\vert \Psi_{\Gamma l\lambda}\rangle$ at a fixed molecular geometry, e. g. the equilibrium position of the negative ion or the equilibrium  of the neutral molecule, which are close to each other. The TDMs then take the simple form
\begin{eqnarray}
\label{eq:vib_tdm_approx}
d_{\pi,\Gamma l\lambda}^{(v\to v_f)}\approx \langle\Psi_f\vert d_\pi\vert \Psi_{\Gamma l\lambda}\rangle_{{\cal Q}_0} \int \chi^*_{v_f}(\vec{q}) \chi_{v}(\vec{q})d\vec{q}  \,,
\end{eqnarray}
where the subscript ${\cal Q}_0$ refers to the equilibrium geometry  of the negative ion. For molecules such as C$_n$H and C$_m$N, for which the potential energy surfaces of the initial electronic state of the target and the final state of the negative ion are quite similar in shape near the equilibrium positions of the ion and the neutral molecule \cite{douguet14}, the Franck-Condon integral in Eq. (\ref{eq:vib_tdm_approx}) is the largest for transitions with $v_f=v$, for which its value is close to unity. For instance, the Franck-Condon integral in Eq. (\ref{eq:vib_tdm_approx}) is about 0.90 for C$_2$H/C$_2$H$^-$ and 0.87 for C$_4$H/C$_4$H$^-$  \cite{douguet14}. For transitions to other vibrational levels, the integral is significantly smaller. Therefore, the DREA cross section is well approximated by 
\begin{equation}
\label{eq:cs_RAE3}
\sigma_{i}\approx\frac{4}{3}\frac{\pi \omega^3 m_e}{k^2\hbar^2 c^3} \sum_{l\pi} \left| \langle\Psi_f\vert d_\pi\vert \Psi_{\Gamma l\lambda}\rangle_{{\cal Q}_0}\right|^2\,,
\end{equation}
where the transition dipole moment is evaluated at the energy of the $M+e^-$ system. The transition dipole moments strongly depend on energy, especially for non-zero partial waves, $l>0$. At low energies, their behavior is described quite well by the Wigner threshold law \cite{wigner48}.  The energy dependence of $\left| \langle\Psi_f\vert d_\pi\vert \Psi_{\Gamma l\lambda}\rangle_{{\cal Q}_0}\right|$ for  C$_2$H/C$_2$H$^-$ and C$_4$H/C$_4$H$^-$ is shown in Figs. \ref{fig:D_E_C2H} and \ref{fig:D_E_C4H}. 

\begin{figure}
\includegraphics[width=7cm]{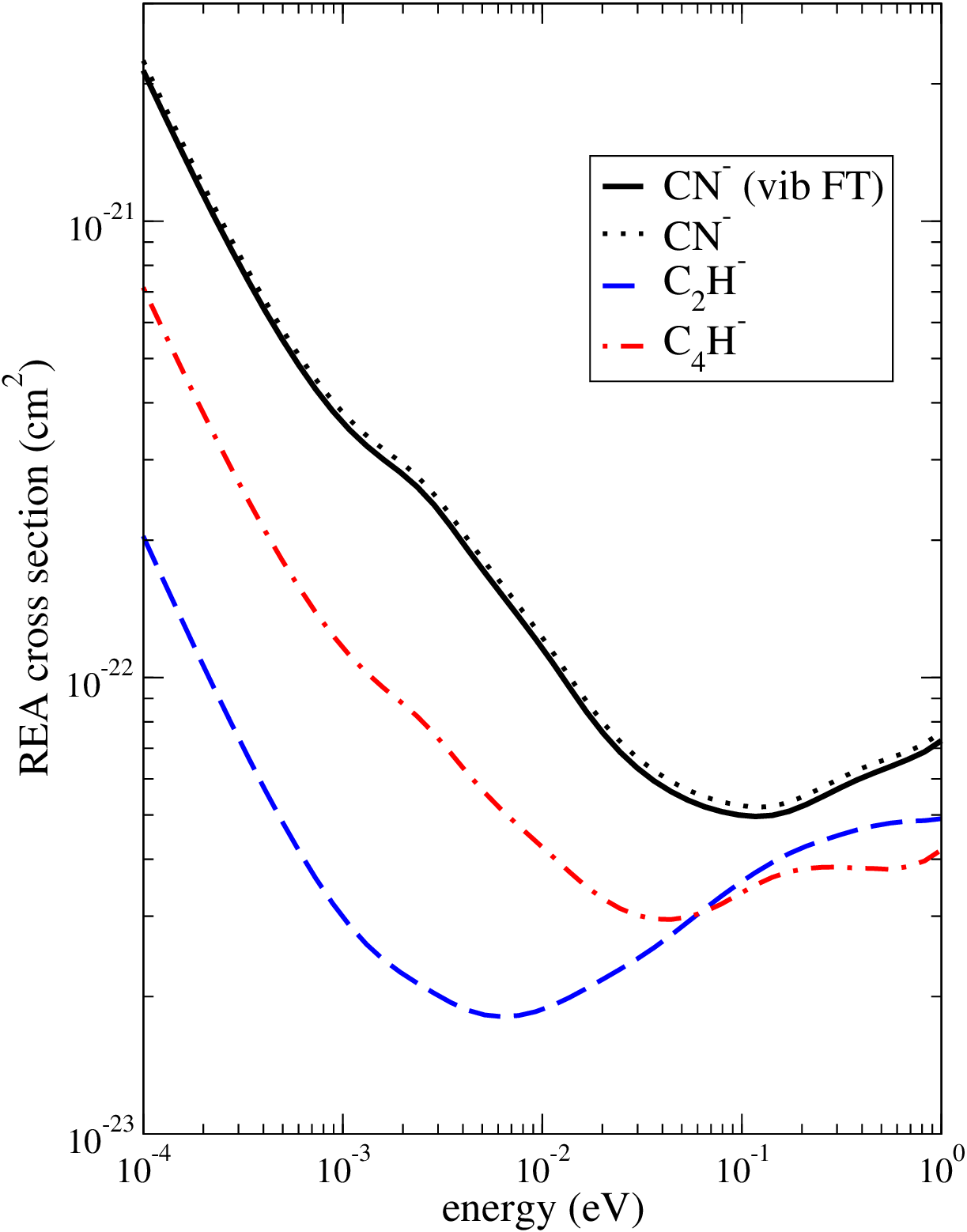}
\caption{The cross sections  for formation of CN$^-$, C$_2$H$^-$, and C$_4$H$^-$ by DREA with the ground vibrational level $v=0$ of the neutral molecule as the initial state. The CN$^-$ cross section is calculated in two different ways:  The dotted line corresponds to the calculation using Eq. (\ref{eq:cs_RAE3}); the cross section shown with the solid line, "vib FT" (vibrational frame transformation), is obtained evaluating the vibrational integral of Eq. (\ref{eq:vib_tdm})  explicitly.}
\label{fig:cs}
\end{figure}

For C$_2$H and C$_4$H, the DREA cross sections were calculated using the approximate formula of Eq. (\ref{eq:cs_RAE3}), with the transition dipole moments determined only at the equilibrium geometry ${\cal Q}_0$ of the negative ion for several energies of the incident electron.  For comparison, the DREA cross section for CN$^-$ has been calculated using the direct integration over internuclear distances, Eqs. (\ref{eq:cs_RAE2}) and (\ref{eq:vib_tdm}), as well as using Eq. (\ref{eq:cs_RAE3}). Figure  \ref{fig:cs} shows the  DREA cross sections obtained for CN$^-$, C$_2$H$^-$, and C$_4$H$^-$. As can be seen in the figure, the CN$^-$ cross sections obtained in the two ways are almost identical. Therefore, Eq. (\ref{eq:cs_RAE3}) gives a very good approximation  of Eqs. (\ref{eq:cs_RAE2}) and (\ref{eq:vib_tdm}) for CN$^-$.  For C$_2$H/C$_2$H$^-$ and C$_4$H/C$_4$H$^-$ the approximation is likely less accurate because the vibrational functions of the neutral molecule and the ion are not as similar to each other as for CN$^-$.

\begin{figure}
\includegraphics[width=8cm]{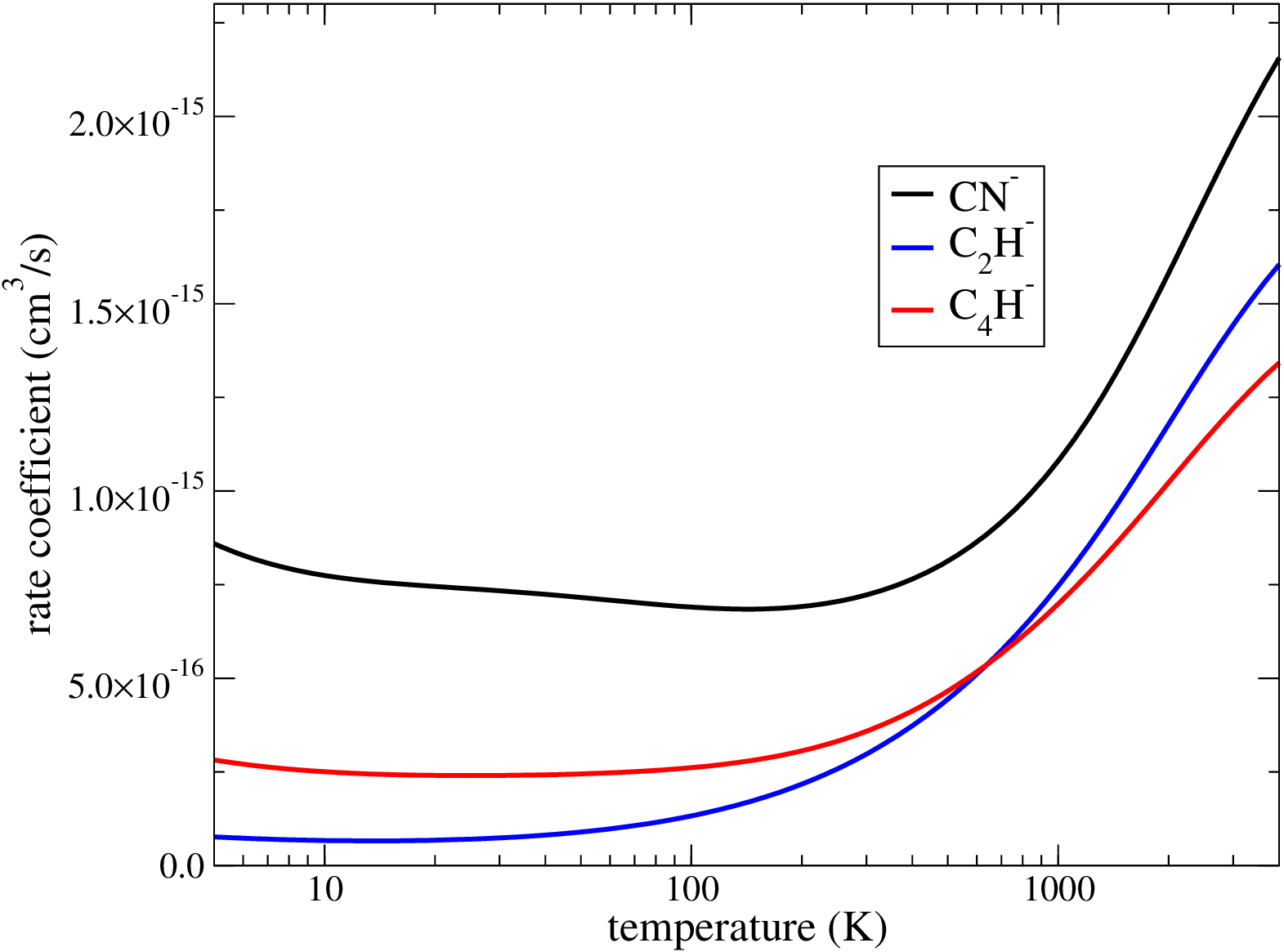}
\caption{The REA thermally-averaged rate coefficients.}
\label{fig:rates}
\end{figure}

The obtained DREA cross sections have been used to determine the thermal rate coefficients, which are shown in Fig. \ref{fig:rates}. The rates for the formations of these three anions via DREA are too small to explain the astrophysical observations. 

\section{Comparison with the results of photodetachment experiments}
\label{sec:photodetachment}

There is no experimental data on radiative electron attachment to the CN, C$_2$H, and C$_4$H molecules. However, the calculated transition dipole moments can be used to determine the photodetachment cross sections, for which experimental data on absolute values of photodetachment cross sections have recently been obtained for CN$^-$ \cite{kumar13}, C$_2$H$^-$ \cite{best11} and  C$_4$H$^-$ \cite{best11} anions. Here, we discuss the interpretation of the photodetachment experiments only briefly. A detailed and more elaborated study can be found in Ref. \cite{douguet14}.
 
The photodetachment cross section  is given by (see, for example, Eq. (2.202) of Ref. \cite{friedrich06})
\begin{equation}
\label{eq:sigma_PD}
 \sigma_{PD}=\frac{4\pi^2 \omega}{3c} \sum_{l\pi} \left| d_{l,\pi,E}\right|^2\,,
\end{equation}
where $\omega$ is the photodetached photon frequency and $d_{l,\pi,E}$ is the dipole moment between the initial bound state and the final continuum state with energy $E$ and partial wave $l$ of the photodetached electron. The radial part of the wave function $\phi_E(r)$ of the initial electronic-continuum state  used to calculate the transition dipole moment $ d_{l,\pi,E}$ in Eq. (\ref{eq:sigma_PD}) is energy-normalized. The continuum functions used in our calculation (see Eq. (\ref{eq:transition-dipole})) are normalized as
\begin{equation}
\phi_E(r)=\sqrt{\frac{2 m_e}{\pi\hbar^2}}\Psi_{\Gamma l\lambda} \,.
\end{equation}
Thus, the PD cross section can be written as
\begin{equation}
\label{eq:cs_PD}
\sigma_{PD}= \frac{8m_e \pi \omega}{3\hbar^2 c}\sum_{l\pi} \left| d_{\pi,\Gamma l-\pi}^{(v\to v_f)}\right|^2\,.
\end{equation}
For convenience of comparison with the results of the DREA calculations, the photon frequency can be expressed in terms of the electron affinity $E_{ea}$ and the energy $E_{el}$ of the incident (in DREA) or emitted (in PD) electron $\hbar\omega=E_{el}+E_{ea}$. This formula assumes that both initial and final vibrational levels are not excited, $v=v^-=0$, and that the zero-point energies for the neutral molecule and the ion are the same. The experimentally measured affinities are 3.862 eV for CN \cite{bradforth93},  2.969 eV for C$_2$H \cite{ervin91,zhou07}, and  3.558 eV for C$_4$H \cite{taylor98}.

\begin{figure}
\includegraphics[width=8cm]{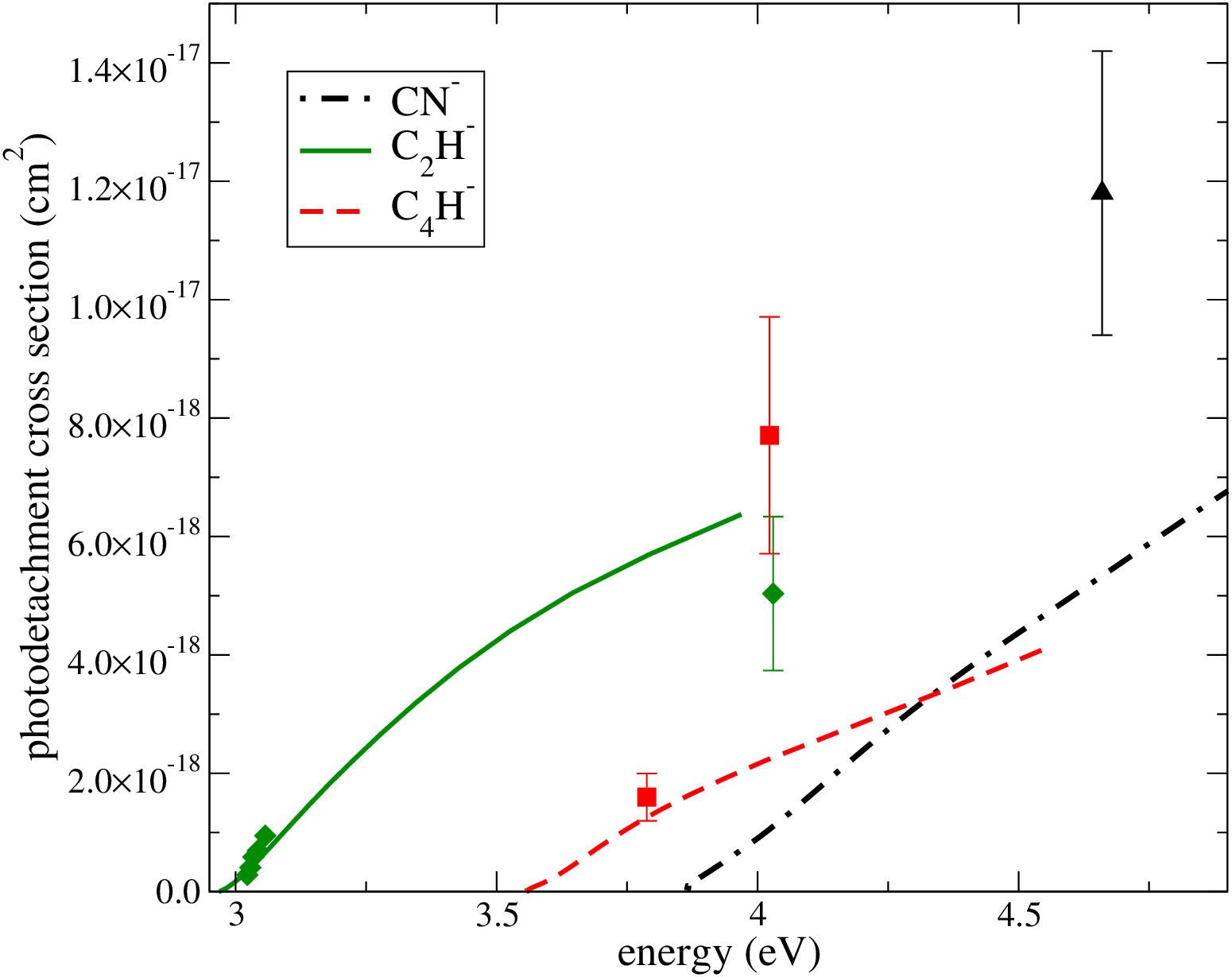}
\caption{Theoretical (lines) and experimental (symbols with uncertainty bars) photodetachment cross sections for CN$^-$, C$_2$H$^-$, and C$_4$H$^-$. The triangle is the experimental result  for CN$^-$ \cite{kumar13}, diamonds  -- C$_2$H$^-$ \cite{best11} , squares  --  C$_4$H$^-$ \cite{best11}.}
\label{fig:cs_PD}
\end{figure}

We discuss now the accuracy of the present calculations. The main source of uncertainty for the REA  and PD cross sections for all three molecules considered here is the quality of electronic continuum and bound state wave functions. The quality of the  electronic bound state wave function could be assessed in part by comparing the obtained theoretical affinity with the experimental one.  The affinities corresponding to wave functions used in the present calculations, are 3.8 eV, 2.2 eV, and 3.0 eV for CN$^-$, C$_2$H$^-$, and C$_4$H$^-$ respectively. Therefore, the agreement is about 1\% for CN$^-$ and much poorer, about 30\%, for C$_2$H$^-$ and C$_4$H$^-$. The quality of  electronic continuum wave functions is more difficult to assess. Based on our previous experience with the electron-scattering calculations, we assume here that an additional uncertainty in the calculated transition dipole moments due to the quality of continuum wave functions is about 10\% for CN$^-$, C$_2$H$^-$, and C$_4$H$^-$. These considerations give an estimated uncertainty in the REA and PD cross sections of about 20\% for  CN$^-$ and 40\% for C$_2$H$^-$ and C$_4$H$^-$. For C$_2$H$^-$ and C$_4$H$^-$, there are additional sources of uncertainty; the neglected geometry dependence of the transition dipole moments and the neglected role of rovibrational Feshbach resonances, which could be present at low collisional energies.

Figure \ref{fig:cs_PD} shows the PD cross sections calculated for CN$^-$, C$_2$H$^-$, and  C$_4$H$^-$ using Eq. (\ref{eq:cs_PD}) and compares them with the available experimental data, which were estimated in Ref. \cite{best11} to have about 25 \% of uncertainty. The agreement is very good for C$_2$H$^-$, especially for experimental data points near the photodetachment threshold. The agreement for C$_4$H$^-$ is also good for the lowest energy point, but is about a factor 2-3 lower than the experimental value for the second energy point. Similarly, the only experimental data point for CN$^-$ measured at 4.65 eV gives a PD cross section twice larger than the theoretical value. The reason for this discrepancy is not clear. Overall, the agreement of the theoretical and experimental results is sufficient to conclude that the theory is reliable for calculations of photodetachment cross sections and, therefore, also for DREA cross sections. Thus, the results of photodetachment experiments validate the present theoretical approach for the DREA process and the results obtained using the approach. In particular, the present results and the PD experiments \cite{kumar13,best11} with C$_4$H$^-$ and CN$^-$ suggest that the observed abundance of these two ions in the ISM can hardly be explained by the DREA mechanism.

\section{The contribution of the indirect process to the total REA cross section: The CN example}
\label{sec:IREA-CN}

We now consider the process of radiative attachment mediated by the non-adiabatic couplings, through the IREA mechanism discussed in the introduction. In this section, we consider the IREA process for CN, for which the non-adiabatic couplings are evaluated numerically. We extend the approach to larger molecules in the next section. 

The probability per unit time of a transition from the initial vibronic state $\vert i\rangle$ of the $e^-+$CN system described above into the final state of CN$^-$ being in a vibrationally excited level $v_f\sim 19$ is given by the Fermi's golden rule
\begin{eqnarray}\label{eq:FGRule}
P={\frac{2\pi}{\hbar}}\left|\langle f | \hat\Lambda | i \rangle\right|^2 \rho(E_c)\,,
\end{eqnarray}
where $\rho(E_c)$ is the density of final states after the electron capture $\rho(E_c)\sim 1/\Delta E_{rv}$, with $\Delta E_{rv}$ the energy splitting between rovibrational states, and $\hat\Lambda$ is the operator of non-adiabatic coupling. Denoting $R$ the CN internuclear distance, the matrix element $\Lambda_{fi}$ between initial and final vibronic states is
\begin{eqnarray}
\label{eq:nac1}
\Lambda_{fi} =\frac{\hbar^2}{\mu_{CN}}\int  \chi^*_{v_f}(R) \Lambda_{f,\Gamma l \lambda}(R) {\frac{\partial}{\partial R}}\chi_{v}(R) dR\,,
\end{eqnarray}
where $\mu_{CN}$ is the reduced mass of CN and $\Lambda_{f,\Gamma l \lambda}(R)$ is the following electronic matrix element 
\begin{eqnarray}
\label{eq:nac2}
\Lambda_{f,\Gamma l \lambda}(R)=\int\Psi^{*}_f(\vec{r},R){\frac{\partial}{\partial R}}\Psi_{\Gamma l\lambda}(\vec{r},R)d\vec{r}\,.
\end{eqnarray}
In the above equation, $\vec {r}$ denotes collectively all electronic coordinates. Note that we neglected the second term of the non-adiabatic couplings, which is expressed with the second derivative of the electronic wave function with respect to the internuclear distance. 

Calculating $\Lambda_{fi}$, we did not account for the integral over rotational coordinates. The latter integral is of the order of unity or smaller. A change of rotational quantum number $j$ during the process of the non-adiabatic electron attachment is unlikely for CN$^-$ at low collision energies. This is because the values of the non-adiabatic couplings are the largest for $s$-wave scattering and much smaller for higher partial waves at low energies. 
Thus, we use the vibrational splitting in the formula for the density of states $\rho(E_c)\sim 1/\Delta E_{v}$.

Numerical evaluation of the electronic matrix element $\Lambda_{f,\Gamma l \lambda}(R)$ of the non-adiabatic couplings was performed at several internuclear distances. Typically, non-adiabatic couplings can be calculated using standard {\it ab initio} programs, but such calculations are limited to couplings between electronic bound states, whereas in the present case, the initial electronic state belongs to the electronic continuum spectrum of the molecule. One possible way to calculate the couplings with a continuum state using standard {\it ab initio} programs  is to include very diffuse orbitals in the bound-state calculations in order to cover the region of large electronic radial distance. As more diffuse functions are added to the basis set, a better description of the asymptotic region is achieved, which leads to the appearance of "box-state" like wave functions with positive asymptotic energy. Such states resemble a scattering state of the electron if enough diffuse functions are added to the basis set. The box states should then be normalized appropriately to represent a continuum state. The latter approach should provide a reasonable approximation of the non-adiabatic couplings because no strong boundary conditions are imposed on the wave function. Performing numerical calculations, we have verified convergence of results (the non-adiabatic couplings) with respect to the number of added diffuse functions.  
 
\begin{figure}
\includegraphics[width=8cm]{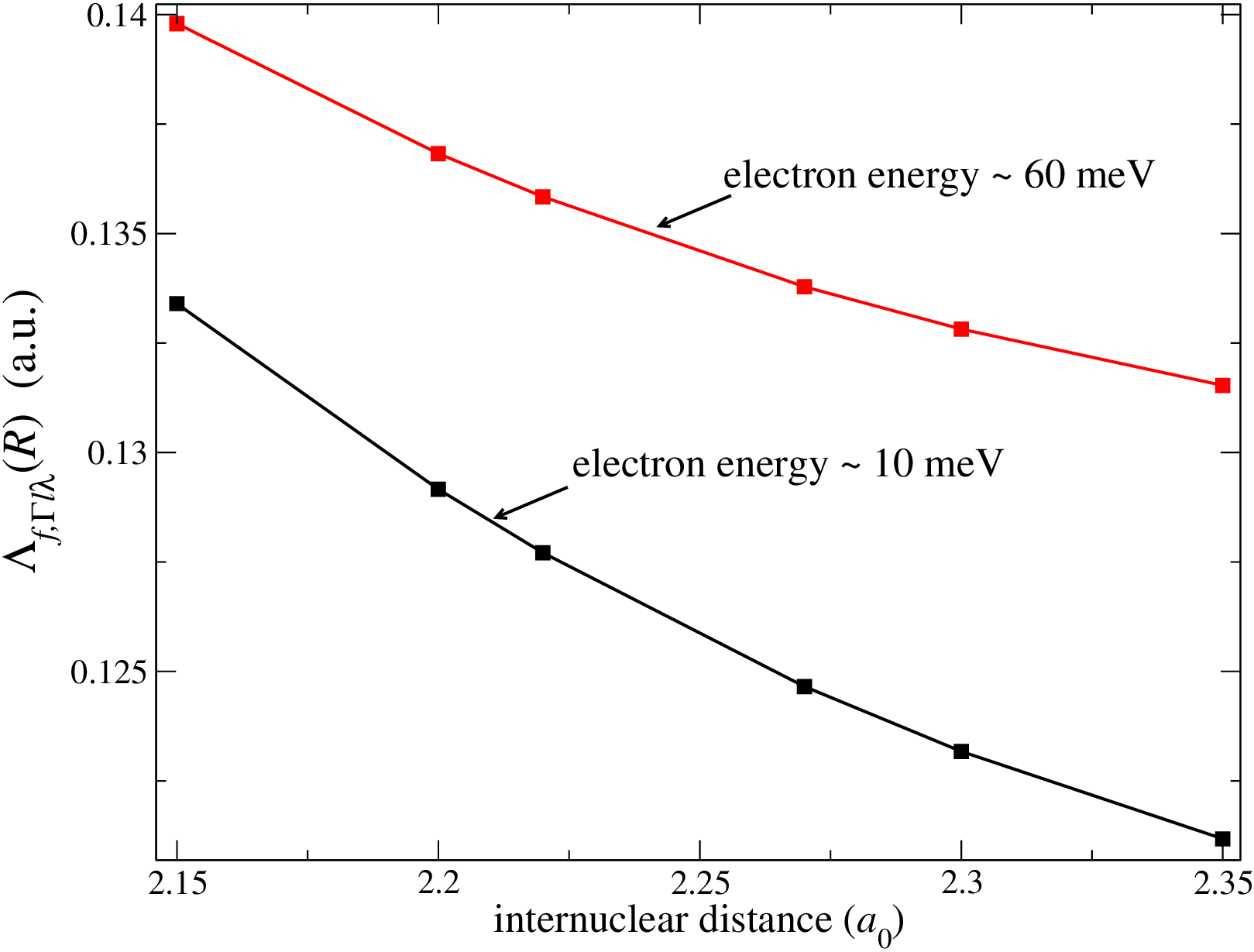}
\caption{The matrix element of the non-adiabatic coupling operator of Eq. (\ref{eq:nac2}) evaluated for two different electronic "quasi-continuum" states with energies $\sim$10  and $\sim$60 meV above the electronic threshold. The continuum state of the incident electron is normalized in the same way as in Ref. \cite{douguet13}, where the current density of electrons in the incident plane wave is equal to the electron velocity (see Eqs. (10), (A2), and (A19) in \cite{douguet13}).}
\label{fig:non_ad_coupling}
\end{figure}

Since the calculations are performed in a finite volume represented by the space spanned by our diffuse basis set, the calculated electronic states of ${\rm CN}+e^-$ just above the CN threshold energy appear as a "quasi-continuum" of states $\Psi^l_n$ with discretized energies. These states are labeled (in order of increasing energy above the electronic threshold) with index $n$ and with the dominant angular momentum $l$. The states should be rescaled to be energy normalized by multiplying by a factor $\Delta E^{-1/2}_{el}$, where $\Delta E_{el}$ is the energy difference between "quasi-continuum" states.  Because $\Delta E_{el}$ is changing from one state to another, it is taken here as the average between two energy differences for three neighboring "quasi-continuum" states. The calculated values of  $\Lambda_{f,\Gamma l \lambda}(R)$ are plotted in Fig. \ref{fig:non_ad_coupling} for two "quasi-continuum" $s$-wave scattering states. These states, corresponding to $n=2$ and $3$, and have respectively $\sim$10  and $\sim$60 meV asymptotic electron energy above the electronic threshold. Notice that the non-adiabatic couplings for the two states are similar and decrease with the internuclear distance. Furthermore, states with higher energies have larger couplings, which is expected from the Wigner threshold law because such couplings should increase approximately as $\sqrt{E_{el}}$. Because the low-energy non-adiabatic couplings associated with $p$-wave scattering states increase only as $E_{el}^{3/2}$, their values at low energy were found to be negligible in comparison with the couplings from $s$-waves scattering states. Finally, although the non-adiabatic couplings were obtained from an approximative treatment, their values should represent an estimation accurate enough for the purpose of this study.  

The electron capture cross section is then obtained by dividing the probability by the current density $j_{cd}$ in the incident wave. If the incident wave is normalized as in Eqs. (A2) and (A19) of Ref. \cite{douguet13}, the current density is then simply equal to the velocity of the incident electron $j_{cd}=v_{cd}$. It gives the following estimation for the cross section
\begin{eqnarray}
\label{eq:cs_irea_flux}
\sigma_c&=&\frac{P}{j_{cd}} \approx \frac{\pi}{E_{el}}\frac{1} {\Delta E_{v}} \frac{\hbar^4}{\mu_{CN}^2}\nonumber\\
&&\times\left|\int \chi_{v_f}^*(R)\Lambda_{f,\Gamma l \lambda}(R) {\frac{\partial}{\partial R}} \chi_{v}(R)dR\right|^2\,.
\end{eqnarray}
In the above equation, we assumed that the ratio of the number of final rotational levels of the formed anion CN$^-$ to the number of initial rotational levels is approximately unity. Moreover, we consider that the value of $\Lambda_{f,\Gamma l \lambda}(R)$ is almost constant in the energy range under study. This represents a relatively good approximation, at least between 10-100 meV, on the view of the weak energy dependence of $\Lambda_{f,\Gamma l \lambda}(R)$, as seen in Fig. \ref{fig:non_ad_coupling}.

We have calculated the vibrational integral numerically with the final vibrational level of CN$^-$  ($v_f=19$), which has an energy close to the energy of the initial vibrational level of CN ($v_i=0$) with a negligible asymptotic energy of the incident electron. We obtained the value
\begin{equation}
\label{eq:pre_c1}
\int \chi^*_{19}(R)\Lambda_{f,\Gamma l \lambda}(R) {\frac{\partial}{\partial R}}\chi_0(R)dR =-1.5 \times 10^{-5} \mathrm{a.u.}\,,
\end{equation}
expressed in a.u. (atomic units). The vibrational energy spacing $\Delta E_{v}$ is about 0.01 hartree (H) and the reduced mass $\mu_{CN}$ is $1.1\times10^4~m_e$ (a.u.).
Therefore, the cross section to capture an electron into a vibrational level is approximately
 \begin{equation}
 \label{eq:est1}
 \sigma_c \approx    \frac{6 \times 10^{-16}}{E_{el}}  a_0^2\,.
\end{equation}

The estimate (\ref{eq:est1}) is an upper bound for the IREA cross section because of the autodetachment process: The  actual cross section is reduced because the formed CN$^-$ ion in the excited vibrational level can decay back to the CN molecule and a free electron. The overall cross section of the indirect REA is then 
 \begin{equation}
 \label{eq:cs_irea}
\sigma_{IREA}=  \frac{\Gamma_{sp}}{\Gamma_{tot}} \sigma_c\le  \sigma_c\,,
\end{equation}
where $\Gamma_{tot}$ is the total width for the decay of the unstable vibrational state of CN$^-$ formed during the collision. For the case of CN/CN$^-$, the total width is the sum $\Gamma_{tot}=\Gamma_{sp}+\Gamma_{ad}$ of the widths towards autodetachment $\Gamma_{ad}$ and towards  spontaneous emission into all possible vibrational levels.

The rate coefficient $\Gamma_{ad}$ can be calculated using the Fermi's golden rule and the same matrix element of the non-adiabatic coupling of Eq. (\ref{eq:nac2}). The density of final states is calculated differently since it will correspond to an outgoing electronic state. In this case, the density of states is unity because the radial part of the wave function is energy-normalized. Therefore, the probability of autodetachment per unit time is roughly
\begin{eqnarray}\label{eq:FGRule2}
\Gamma_{ad}\sim{\frac{2\pi}{\hbar}}|\Lambda_{fi}|^2 \sim   10^{-17}\ \mathrm{H}/\hbar=0.41\ \mathrm{s}^{-1} \,,
\end{eqnarray}
if, at the end of the process, the CN molecule is again in the ground vibrational level.

The rate $\Gamma_{sp}$ of spontaneous emission can also be estimated using the standard formula (see Eq. (2.189) of Ref. \cite{friedrich06})
 \begin{equation}
\Gamma_{sp}\sim\frac{\omega_{sp}^3}{\hbar c^3}\sum_{\pi} \left| d_{\pi}^{(v^-\to {v'}^-)}\right|^2\,,
 \end{equation}
where $d_{\pi}^{(v^-\to {v'}^-)}$ is the vibrational matrix element of the {\it permanent} dipole moment of the anion $d_{\pi}(R)$ for vibrational transition  $v^-\to {v'}^-$~:
 \begin{eqnarray}
\label{eq:per_dipole}
d_{\pi}^{(v^-\to {v'}^-)} =\int  \chi_{{v'}^-}^*(R) d_{\pi}(R) \chi_{v^-}(R) dR\,.
\end{eqnarray}
 In the above equation ${v'}^-$ is the vibrational level of CN$^-$ after emission of a photon, which should be smaller than $v^-$ in order to stabilize the anion. The vibrational matrix element are the largest for $\Delta v^{-}=1$.  For such a transition, the vibrational dipole matrix element is roughly equal to the derivative of $d_{\pi}(R)$ with respect to $R$. The order of magnitude of the derivative is about unity in a.u. and, therefore, the vibrational matrix element of the permanent dipole moment is about $ea_0$. For transitions with $\Delta v^{-}>1$, the matrix elements are significantly smaller. Thus, we only account  for the ${v'}^-={v}^- -1$ transition, for which $\omega_{sp}=\Delta E_v/\hbar$, and obtain $\Gamma_{sp}\sim 1.3\times 10^{-17}$ H$/\hbar=0.54\ \mathrm{s}^{-1}$. The two rate coefficients  $\Gamma_{ad}$ and $\Gamma_{sp}$ are thus comparable to each other.

\section{The indirect REA in larger molecules}
\label{sec:IREA-larger}
For large polyatomic molecules, one expects that the electron capture into excited vibrational levels of the negative ion $M^-$ should be more efficient than for the CN$^-$ molecule. Nevertheless, the electronic non-adiabatic couplings along a single coordinate of $M^-$ are not expected to be much different than in CN$^-$. The latter statement should be true for molecules that do not exhibit singular effects near threshold, e.g. effects due to the presence of a resonance or virtual state, or if the radical ground state has an unusually large permanent dipole moment. In order to obtain another basis for comparison, we have also calculated the non-adiabatic couplings in the case of C$_2\rm{H}^-$ and found only slightly smaller values than in CN$^-$. On the other hand, for molecules with large permanent dipole (e.g. C$_6$H, C$_8$H or C$_3$N), threshold effects could be important and the non-adiabatic couplings significantly larger. The study of the role of a permanent dipole moment in REA is out of the scope of the present study and has been discussed elsewhere \cite{guthe01,carelli14}. 

Therefore, in the following development, we only consider the role of the number of degrees of freedom in the electron capture probability and discard the possible role of unusual threshold effects or vibrational Feshbach resonances. Because the number of non-Born-Oppenheimer couplings and the energy-density $\rho_M(E_c)$ of available vibrational levels of $M^-$ rises with the number of degrees of freedom, the overall capture rate is expected to increase correspondingly. One can readily estimate the density of final states in a molecule with several degrees of freedom. Let us thus consider a given linear molecular anion $M^-$ and radical $M$ with $\cal S$ degrees of freedom. Henceforth, the molecular ion $M^-$ and its counterpart radical $M$ will be represented as a set of vibrational harmonic oscillators with respective normal coordinates $q_s$ and $q_s'$, and associated harmonic vibrational frequencies $\omega_s$ and $\omega_s'$. Both sets of normal coordinates represent the displacements around the equilibrium positions of the anion and neutral molecule, respectively. It will be shown later that for large enough carbon chain molecules, the harmonic oscillator model represents an accurate approximation. More importantly, we will not lose generality of the results by using the harmonic oscillator approximation in our model. This is a good approximation near the minima of the  $M$ and $M^-$ potential energy surfaces. 

In order to obtain a rough estimation of the density of vibrational levels, we first introduce a characteristic energy splitting $\Delta E_c$ for the different vibrational levels. We then assume that the energy splitting in different modes $q_s$ is approximately equal to the average energy splitting, namely $\hbar\omega_s\approx\Delta E_c$ for all $s$ (this approximation will be relaxed later). If we denote by $v_s$ the number of excited quanta along the $q_s$ coordinate, then the total number of excited quanta ${\cal V}_0$ corresponding to the energy of the system just above the radical threshold will be approximately given by:
\begin{equation}
\label{eq:v_m}
 {\cal V}_0\equiv\sum_{s=1}^{\cal S} v_s\approx\frac{E_{ea}}{\Delta E_c}\,.
\end{equation}
The affinity $E_{ea}$ of the interstellar anions ranges from about $3-4$ eV, while the vibrational frequencies are usually of the order of $100-3000$ cm$^{-1}$. For the frequencies of the stretching modes, which are commonly the largest, this gives approximately $10<{\cal V}_0<30$. On the other hand, the frequencies of the bending modes are small ($\hbar\omega_s<800$ cm$^{-1}$) and ${\cal V}_0$ could grow as large as ${\cal V}_0> 300$ for very loose bending modes.
For a given value of ${\cal V}_0$, the number of vibrational levels of $M^-$ in the interval of energy $\Delta E_c$ is approximately given by (see Appendix \ref{app:A})
\begin{equation}
\label{eq:N_T_text}
N_T={{{\cal V+S}-1}\choose{\cal V}}\,.
\end{equation}
Therefore, the density of vibrational levels is simply:
\begin{equation}
\rho_M(E_c)\sim \frac{1}{\Delta E_c}{{{\cal V+S}-1}\choose{{\cal V}}}\,.
\end{equation}
The density of levels grows rapidly with the number of degrees of freedom and the total number of quanta. The largest contribution to the density comes from excitation of several modes at once, as shown in appendix \ref{app:B}. 

\subsection{Model to evaluate the IREA cross section}
Let us now introduce the model describing the IREA mechanism to large carbon chain radicals, neglecting the rotational motion. We need to evaluate the non-adiabatic couplings between the initial state of the $M + e^-$ system and a vibronic state of $M^-$ into which the electron is captured. It is convenient to introduce the dimensionless normal coordinates $\bar{q}_s=q_s\sqrt{\mu_s\omega_s/\hbar}$ of the negative ion $M^-$, where $\mu_s$ is the reduced mass. In a similar way, dimensionless  coordinates $\bar{q_s}'$ are also introduced for the neutral molecule $M$. For the molecules under consideration, the potential energy surfaces of $M$ and $M^-$ are nearly parallel, such that the normal coordinates $\bar{q_s}$ and $\bar{q_s}'$, as well as the vibrational frequencies $\omega_s$ and $\omega_s'$, are almost identical for all modes. This is the reason why we will use the same notations $\bar{q_s}$ for the normal coordinates of $M$ and $M^-$. The only difference between the normal coordinates of $M$ and $M^-$, which will be accounted for below in evaluating the Franck-Condon overlaps, is the displacement between the equilibrium positions of $M$ and $M^-$. Note that this approximation is not only convenient, but has been shown numerically to be excellent for CN \cite{douguet13}, as well as for the hydrocarbon chains C$_2$H, C$_4$H and C$_6$H \cite{douguet14}.

The non-Born-Oppenheimer operator $\hat{\Lambda}^{(v)}$, acting on the nuclei coordinates and integrated over the electronic degrees of freedom, takes the form
\begin{eqnarray}
\label{eq:nac4}
\hat{\Lambda}^{(v)}\equiv\langle\Psi_f|\hat{\Lambda}|\Psi_{\Gamma l\lambda}\rangle_{\vec r}=\sum_{s=1}^{{\cal S}} \Lambda_s(\bar{q}_s) {\frac{\partial}{\partial\bar{q}_s}}\,,
\end{eqnarray}
where the electronic coupling $\Lambda_s(\bar{q}_s)$ (equivalent to the operator introduced in Eq. (\ref{eq:nac2}) for the one-dimensional case) has the following value
\begin{eqnarray}
\label{eq:nac5}
\Lambda_s(\bar{q}_s)=\int\Psi^{*}_f(\vec{r},\vec{q}){\frac{\partial}{\partial \bar{q}_s}}\Psi_{\Gamma l\lambda}(\vec{r},\vec q)d\vec{r}\,.
\end{eqnarray}

We now assume that  $\Lambda_s(\bar{q}_s)$ are almost constant over the vibrational displacements, hence $\Lambda_s(\bar{q}_s)\approx\Lambda_s$. This approximation has shown to be excellent in the case of CN$^-$ and C$_2$H$^-$, for which the electronic non-Born-Oppenheimer matrix element does not vary significantly over large inter-nuclei displacements (see Fig. \ref{fig:non_ad_coupling}). We also assume that all the couplings have about the same value along the different modes, namely $\Lambda_s(\bar{q}_s)\approx\Lambda_0$ for all $s$. With these approximations, the operator in Eq. (\ref{eq:nac4}) becomes simply
\begin{eqnarray}
\label{eq:nac3}
\hat{\Lambda}^{(v)}=\Lambda_0\sum_{s=1}^{{\cal S}}{\frac{\partial}{\partial\bar{q}_s}}\,.
\end{eqnarray}
The calculation of the matrix elements of $\hat{\Lambda}^{(v)}$ requires the evaluation of overlaps between the initial and final vibrational wave functions, which cannot be expressed in analytical form in a general case.  

The matrix element of the non-Born-Oppenheimer operator (\ref{eq:nac3}) to be evaluated is given by
\begin{eqnarray}
\label{eq:nac_multi}
\Lambda_{fi}=\langle\chi_f|\hat{\Lambda}^{(v)}|\chi_i\rangle\,,
\end{eqnarray}
where $|\chi_i\rangle=|00\cdots0\rangle$ denotes the vibrational ground state of $M$ and $|\chi_f\rangle=|v_1v_2\cdots v_{{\cal S}}\rangle$ denotes an excited vibrational resonant state of $M^-$. The vibrational wave functions are expressed as a product of harmonic functions of the $\bar{q}_s$ displacements with $v_s$ quanta in each normal mode. 
If we now introduce the rescaled non-Born-Oppenheimer elements $\bar{\Lambda}_{fi}=\Lambda_{fi}/\Lambda_{o}$, independent of the magnitude of the electronic non-adiabatic coupling, and insert Eq. (\ref{eq:nac3}) into Eq. (\ref{eq:nac_multi}), we obtain
\begin{eqnarray}
\label{eq:nac_multi2}
\bar{\Lambda}_{fi}= \sum_{s=1}^{{\cal S}}\langle v_1|0\rangle\cdots\langle v_s|\frac{\partial}{\partial \bar{q}_s}|0\rangle\cdots\langle v_{\cal S}|0\rangle.
\end{eqnarray}
In the above sum, we call the mode $s$ with the derivative $\partial/\partial \bar{q}_s$ as the \textit{active}  mode, whereas all other modes will be referred as \textit{passive}.
In Eq. (\ref{eq:nac_multi2}), the one-dimensional harmonic vibrational overlaps for passive modes take a simple form. If we introduce the coefficients $\beta_s=\Delta \bar{q}_s/\sqrt{2}$, where $\Delta \bar{q}_s$ denotes the separation between the equilibrium geometries of $M^-$ and $M$ along the dimensionless normal coordinate $\bar{q}_s$, then the vibrational overlap is simply given by \cite{frank99}
 \begin{eqnarray}
\label{eq:frank}
\langle v_s|0\rangle=\frac{\beta_s^{v_s}}{\sqrt{v_s!}}e^{-\frac{\beta_s^2}{2}}\,.
\end{eqnarray}
We would like to stress the following points:
\begin{itemize}
\item The value of $\beta_s$ accounts for the frequency, reduced mass, and length displacement of each mode. In terms of the original (not dimensionless) normal coordinates $\Delta q_s$, these coefficients are $\beta_s=\Delta q_s\sqrt{\mu_s\omega_s/2\hbar}$  \cite{frank99}.
\item The bending modes have $\beta_s=0$ because no displacement exists between the minimum of the surface potentials of $M^-$ and $M$ along these modes. Therefore, the bending modes will only contribute negligibly in the electron capture.
\item The sum over all vibrational quanta of the squared of the overlaps is unity, as required.
\end{itemize}

The overlap, involving an active mode, is obtained by expressing the derivative operator using raising and lowering operators
\begin{eqnarray}
\label{eq:nac_d/dq}
\langle v_s|\frac{\partial}{\partial \bar{q}_s}|0\rangle=\frac{\beta_s^{v_s-1}}{\sqrt{v_s!}}\left(v_s-\beta_s^2\right)\frac{e^{-\frac{\beta_s^2}{2}}}{\sqrt{2}}\,.
\end{eqnarray}
In the case of a bending mode ($\beta_s=0$) the expression (\ref{eq:nac_d/dq}) vanishes for all $v_s$ except for $v_s=1$. It means that  a degenerate active mode can capture an electron through non-Born-Oppenheimer coupling by an excitation of only a single  quantum of vibrational excitation.

The capture probability per unit time into the vibrational state $|\chi_f\rangle$ of the anion is proportional to the square of the element of the non-Born-Oppenheimer operator in Eq. (\ref{eq:FGRule}):
\begin{eqnarray}
\label{eq:lambda}
|\bar{\Lambda}_{fi}|^2=\sum_{s=1}^{\cal S}\frac{\beta_1^{2v_1}\cdots\beta_s^{2v_s-2}\cdots\beta_{{\cal S}}^{2v_{{\cal S}}}}{v_1!\cdots v_s!\cdots v_{{\cal S}}!}\left(\beta_s^2-v_s\right)^2\frac{e^{-\beta_0^2}}{2}\,.\nonumber
\end{eqnarray}
The coefficient $\beta_0$ in the above equation represents the dimensionless displacement between the minima of the neutral and anion potentials in the multi-dimensional space spanned by the $\cal S$ normal coordinates. It is simply given by $\beta_0=\sqrt{\beta_1^2+\cdots+\beta_s^2+\cdots+\beta^2_{{\cal S}}}$~. 

In our treatment, we did not calculate the positions and widths of the vibrational resonances, but instead computed the capture probability in each "quasi-bound" vibrational state of $M^-$. For this reason, we now calculate an electron capture probability per unit time $\langle P\rangle$, averaged over a certain electronic energy interval ${\cal I}_{\Delta}=[E_{tot}-\Delta/2,E_{tot}+\Delta/2]$ around the total energy of the system. Then, the total capture probability per unit time is given by summing over all vibrational states whose energies are situated within the interval ${\cal I}_{\Delta}$, namely
\begin{eqnarray}
\label{eq:P_total}
\langle P\rangle=\frac{2\pi}{\hbar}\frac{\Lambda_0^2}{\Delta}\sum_{f\in{\cal I}_\Delta}|\bar{\Lambda}_{fi}|^2,
\end{eqnarray}
where $|\bar{\Lambda}_{fi}|^2$ is given by Eq. (\ref{eq:lambda}). In this approach, the probability in Eq. (\ref{eq:P_total}) has to be evaluated numerically to obtain an exact value. However, $\langle P\rangle$ can be estimated analytically using the following arguments. First, we fix the total number of quanta ${\cal V}_0$ introduced in Eq. (\ref{eq:v_m}), which is associated with an averaged total energy $\langle E\rangle$. Then, we evaluate the sum of transition probabilities to any vibrational states of $M^-$ with ${\cal V}_0$ excited quanta. Denoting ${\cal V}_f=v_1+\cdots+v_{\cal S}$ the total number of quanta of a vibrational state $|\chi_f\rangle$ and introducing the sum
\begin{eqnarray}
\label{eq:sum}
{\mathcal I}({\mathcal V}_0)=\sum_{ {\cal V}_f={\cal V}_0}|\bar{\Lambda}_{fi}|^2,
\end{eqnarray}
which is similar to the sum in Eq. (\ref{eq:P_total}), but now evaluated fixing the number of quanta instead of the energy interval, then the capture probability in Eq. (\ref{eq:P_total}) can be approximated by the probability per unit time of exciting ${\cal V}_0$ quanta, namely
\begin{eqnarray}
\label{eq:P_total2}
\langle P({\cal V}_0)\rangle=\frac{2\pi}{\hbar}\frac{\Lambda_0^2}{\Delta E_c}{\mathcal I}({\mathcal V}_0).
\end{eqnarray}
In the above equation, $\Delta E_c$ is a characteristic energy spacing, taken as the difference between the average energies corresponding to ${\cal V}_0$ and ${\cal V}_0+1 $ excited quanta. The formula for ${\mathcal I}({\mathcal V}_0)$ is derived in Appendix \ref{app:2},
\begin{eqnarray}
\label{eq:I(V_0)}
{\mathcal I}({\cal V}_0)\approx\frac{\beta_0^{2{\cal V}_0-2}({\cal V}_0+{\cal S}-1)}{({\cal V}_0-1)!}\frac{e^{-\beta_0^2}}{2}\,.
\end{eqnarray}
Choosing the vibrational ground state of $M^-$ as the reference energy, the average energy weighted over the capture probability to excite ${\cal V}_0>1$ quanta, is given by
\begin{eqnarray}
\label{eq:average_energy}
\langle E\rangle\approx({\cal V}_0-1)\hbar\langle\omega\rangle,
\end{eqnarray}
where $\langle\omega\rangle$ is an averaged vibrational frequency (see Appendix \ref{app:2} for details). Therefore, the averaged cross section for electron capture in the IREA process of polyatomic molecules is written in this model as 
\begin{eqnarray}
\label{eq:cs_IREA_averaged}
\langle\sigma_c\rangle&=&\frac{\pi}{E_{el}}\frac{\Lambda_0^2}{\Delta E_c}{\mathcal I}({\mathcal V}_0)\nonumber\\
&=&\frac{\pi}{2E_{el}}\frac{\Lambda_0^2}{\hbar\langle\omega\rangle}\frac{({\cal V}_0+{\cal S}-1)}{({\cal V}_0-1)!}\beta_0^{2{\cal V}_0-2}\ e^{-\beta_0^2}\,,
\end{eqnarray}
where the current density in the incident plane wave of electrons, as well as the normalization of radial wave function, are the same as in Eq. (\ref{eq:cs_irea_flux}).

\begin{figure}[t]
\includegraphics[width=8cm]{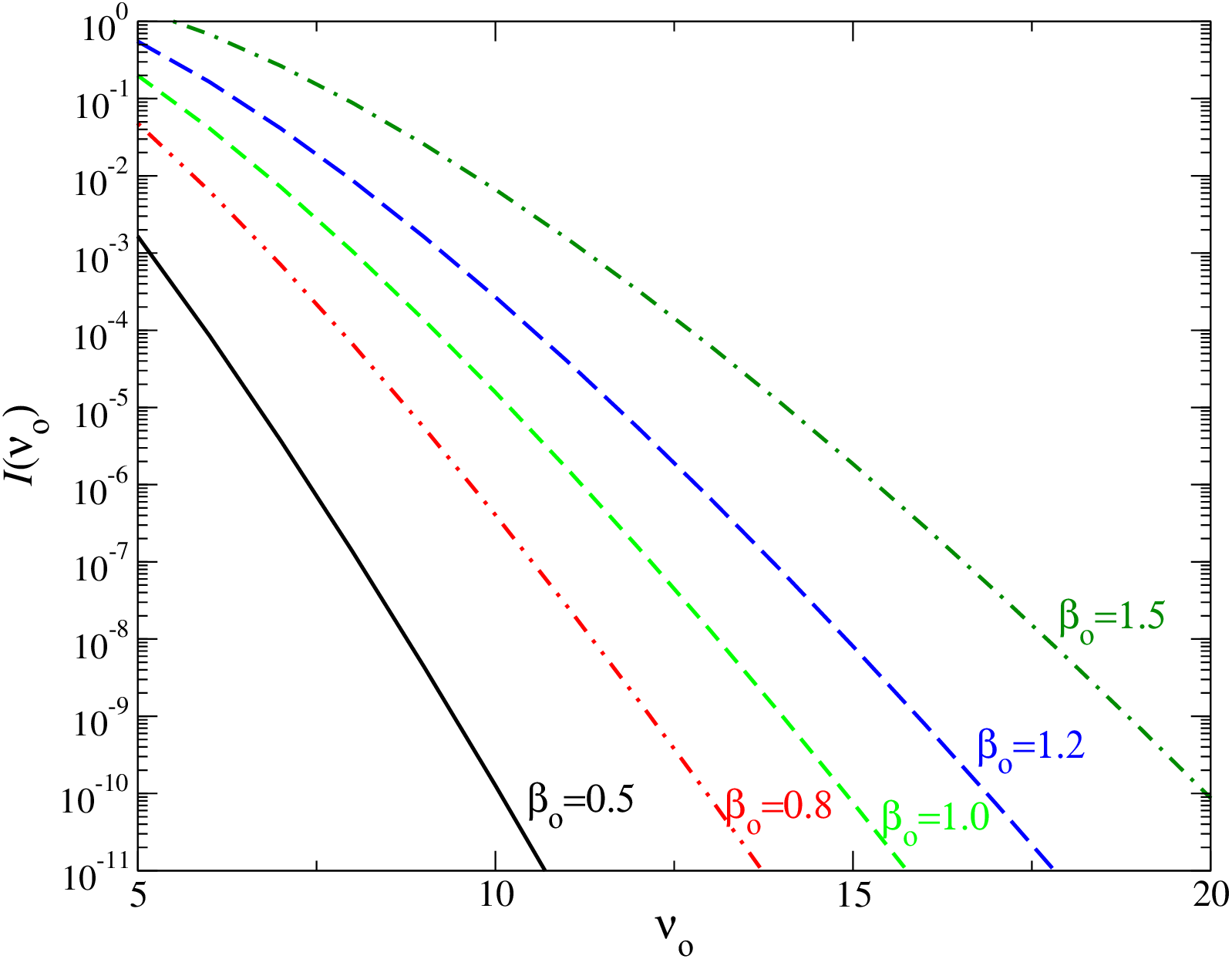}
\caption{Values of ${\mathcal I}({\cal V}_0)$ for $5\leq{\cal V}_0\leq20$ and ${\cal S}=22$.}
\label{fig:beta}
\end{figure}

The parameters entering Eq. (\ref{eq:cs_IREA_averaged}), namely the equilibrium positions, normal modes, average frequency and shift $\beta_0$, should be calculated separately for each molecule. Below, we give typical values of these parameters for the molecules under study. For bending modes in linear molecules, the shifts $\beta_s$ between equilibrium positions of $M$ and $M^-$ are equal to zero. Therefore, only stretching modes contribute into $\beta_0$.  The average frequency of the stretching modes in large carbon chains is $\langle\omega\rangle\approx 2000~\rm{cm}^{-1}$.  Using a typical value of 4 eV for electronic affinity of carbon chain molecules C$_{2n}$H and C$_{2n-1}$N ($n$ is an integer) and assuming a small energy of the incident electron, we obtain $13<{\cal V}_0<16$. Our calculated values of  $\beta_0$ are 0.19 (CN), 0.46 (C$_2$H), 0.56 (C$_4$H) and 0.39 (C$_6$H) \cite{douguet14}. The latter value, corresponding to the $^2\Pi$ ground state of C$_6$H, was calculated at the Hartree-Fock level, and could be somewhat underestimated. In Fig. \ref{fig:beta}, the values of ${\mathcal I}({\cal V}_0)$ are plotted as a function of ${\cal V}_0$ for different values of $\beta_0$ and choosing, for instance, the number of degrees of freedom in C$_8$H, namely  ${\cal S}=22$. As evident from the figure, ${\mathcal I}({\cal V}_0)$ is very sensitive to $\beta_0$ but remains small for the typical $\beta_0$ values. As a result, the IREA cross section is small, unless $\beta_0$ is much larger than unity, which is rather unlikely for other carbon chain anions that  we have not studied. 

Taking the largest value of $\beta_0=0.56$ obtained for C$_4$H/C$_4$H$^-$ with $\cal S$=10, $\langle\omega\rangle=0.01$~H, ${\cal V}_0=15$ and $\Lambda_0=0.13$, which is similar to the one obtained for CN/CN$^-$ (see Fig. \ref{fig:non_ad_coupling}), we obtain
\begin{equation}
 \langle\sigma_c\rangle \approx    \frac{5\times10^{-17}}{E_{el}}  a_0^2\,,
\end{equation}
where the energy is in hartree (H). At electron energy of 1~meV, the above formula gives the cross section of $10^{-12}\ a_0^2$. For somewhat more favorable situation, when $\cal S$=22 (as for C$_8$H) and a significantly larger  $\beta_0=1$ (an arbitrary value, not necessarily representing C$_8$H/C$_8$H$^-$), the cross section at the same energy becomes larger, about $10^{-5}\ a_0^2$, but still remains small to explain the astrophysical observations.


\subsection{Statistical distribution in the model}

We consider here the distribution of different anion vibrational states after the electron capture if ${\cal V}_0$ is fixed. The electron capture probability into a vibrational state of the negative ion is proportional to the non-adiabatic coupling elements in Eq. (\ref{eq:lambda}). In this expression,  $\beta_s^2\ll v_s$ in the factor $(\beta_s^2-v_s)$ and, therefore, $\beta_s^2$ can be neglected. Making the change of variable $v_s\to v_s+1$ for the active mode in Eq. (\ref{eq:lambda}) and summing over all the modes, we obtain 
\begin{equation}
|\bar{\Lambda}_{fi}|^2\approx({\cal V}_0+{\cal S}-1)\frac{\beta_1^{2v_1}\cdots\beta_s^{2v_s}\cdots\beta_{{\cal S}}^{2v_{{\cal S}}}}{v_1!\cdots v_s!\cdots v_{{\cal S}}!},
\end{equation}
with the new condition that $v_1+\cdots+v_{\cal S}={\cal V}_0-1$. The most probable configuration \{$v^0_1,\cdots v^0_{\cal S}$\} to capture an electron corresponds to an extremum of the function
\begin{eqnarray}
{\cal G}(v_1\cdots v_{{\cal S}})=\ln{|\bar{\Lambda}_{fi}|^2}-\lambda(\sum_{s=1}^{{\cal S}}v_s-{\cal V}_0+1),
\end{eqnarray}
where we use the method of Lagrange multipliers. Using the  Stirling approximation,
\begin{eqnarray}
\ln(v_s!)\approx v_s\ln\left(\frac{v_s}{e}\right)+\frac{1}{2}\ln(v_s)+\ln(\sqrt{2\pi})\,,
\end{eqnarray}
the Lagrange equations, giving the extremum, become
\begin{eqnarray}
\label{eq:lagrange}
\ln\beta_s^2-\ln v^0_s+\frac{1}{2v^0_s}-\lambda=0.
\end{eqnarray}
Typically, $\beta^2_s\ll1$ and $v^0_s\ge2$, thus the third term in (\ref{eq:lagrange}) is much smaller than the first two terms and can, therefore, be discarded. Solving for $v^0$ we obtain
\begin{eqnarray}
v^0_s\approx\left[\frac{\beta_s^2}{\beta_0^2}({\cal V}_0-1)\right],
\end{eqnarray}
where the brackets denote the closest integer number because  $v^0_s$ should be an integer number. The modes with the largest separation between the equilibrium positions are, therefore, the most strongly excited in the electron capture, whereas bending modes barely contribute. However, if we consider the ${\cal S}_{str}=N-1$ stretching modes of a $N$-atoms carbon chain have similar displacements $\beta_s$, then the stretching modes will be equally populated with $\left[({\mathcal V}_0-1)/{\cal S}_{str}\right]$ quanta. Since we are interested in large carbon chains ${\cal S}_{str}\ge 4$, the optimum occupation numbers are such that $v_s^0\le4$. This result shows why the harmonic oscillator model is accurate for a carbon-chain molecule with several degrees of freedom. States with smallest possible number of excited quanta along each mode but involving many modes, contribute the most into the capture probability. Finally, only vibrational states with energies within the energy interval  $[\langle E\rangle-\Delta E;\langle E\rangle+\Delta E]$ with $\Delta E=\hbar\Delta \omega\sqrt{{\cal V}_0-1}$ around energy $\langle E\rangle$ contribute significantly in the electron capture probability (see Appendix \ref{app:2}). Because $\Delta \omega\approx 1000$~cm$^{-1}$, the energy spread is of the same order as the characteristic energy spacing $\Delta E_c$, such that the cross section of Eq. (\ref{eq:cs_IREA_averaged}) should give a correct estimate of the electron capture cross section. Nonetheless, because the affinity of a negative ion can usually not be assigned exactly to one  ${\cal V}_0$ value,  Eq. (\ref{eq:cs_IREA_averaged}) only gives an order of magnitude of the cross section.

\subsection{Competition between radiative stabilization and autodetachment}
For large molecules, once the electron is captured, in addition to the processes of autodetachment and spontaneous emission from the formed resonance, the system can also change its vibrational state. Although the energy of the system cannot change without emitting a photon, such a change in the vibrational excitation is possible because all vibrational levels in this energy region are in fact resonances and have finite widths. In large polyatomic molecules the coupling between vibrational modes could be strong, so one would expect that, at least, for some molecules a rapid change of the vibrational state is possible. If it happens that the autodetachment width of this new state is small or the probability of the spontaneous emission for this state is enhanced, for example, due to a better Franck-Condon overlap, one can expect that the system will not be able to loose rapidly the attached electron and will eventually be stabilized by emitting a photon. If this happens, then the IREA cross section  (\ref{eq:cs_irea}) should be modified as
 \begin{equation}
 \label{eq:cs_irea2}
\sigma_{IREA}=  \frac{\Gamma_{sp}+\Gamma_{vc}}{\Gamma_{tot}} \sigma_c\,,
\end{equation}
where $\Gamma_{tot}$ now includes also the width $\Gamma_{vc}$ for the mentioned change of the vibrational state $\Gamma_{tot}=\Gamma_{sp}+\Gamma_{ad}+\Gamma_{vc}$. The time scale for change of vibrational  state in a favorable case when the initial and final vibrational states strongly interact with each other, must be of the order of a vibrational period, in a ps range, which makes $\Gamma_{vc}$ dominant over the two other widths. In this limiting case, the IREA cross section $\sigma_{IREA}$ would be determined by its upper bound $ \sigma_c$ of Eq. (\ref{eq:est1}). Based on the above discussion, we can conclude that in the process of formation of negative ions C$_m$N$^-$ ($m=1,3,5$) and C$_n$H$^-$ ($n=2,4$) by radiative attachment, a significant contribution into the total REA cross section of the IREA is unlikely. 

Non-adiabatic couplings in polyatomic molecules could be significant near a conical intersection of potential energy surfaces and, as a result, lead to a larger probability of electron capture during the first step of IREA. For the considered case of radiative attachment, this can happen for certain incident electron energies if (a) the neutral molecule has degenerate vibrational modes and (b) the formed electronic state of the anion is also degenerate. The linear carbon chain molecules C$_n$H ($n=2,4,6,8$) and C$_m$H ($m=3,5$) do have degenerate vibrational modes, and some of the corresponding anions may have excited electronic states of the degenerate $^1\Pi$ symmetry. If such excited electronic states have appropriate energies, an electron with $\pi$ symmetry, incident at the neutral target, could be captured due  non-adiabatic Renner-Teller coupling, into the degenerate $^1\Pi$ state of the anion, exciting at the same time, the degenerate vibrational mode of the molecule.  Such a process is similar to electron capture in dissociative recombination of linear molecules, HCO$^+$ and N$_2$H$^+$ \cite{mikhailov06,douguet08b,douguet09,samantha14}, but there is an important difference: In collisions between an electron and a positive molecular ion, the density of electronic resonances is larger by several orders of magnitude. In dissociative recombination, electronic Rydberg resonances with a vibrationally excited core could be found virtually at any energy above the lowest ionization limit. But only a few molecular anions, such as C$_2^-$, are known to have electronic resonances. Therefore, for a significant increase of the IREA cross section, the anion should have an electronic  $^1\Pi$ state close to the energy of the initial continuum state of the $e^--M$ system. This could, probably, be the case for one of the carbon chain anions, but it seems unlikely that all observed anions have an electronic resonance near the energy of the initial  state of the $e^--M$ system  with small energy of the incident electron.

\section{Conclusion}
\label{sec:conclusion}

We have extended the theoretical approach to study the process of radiative electron attachment, developed in our previous study \cite{samantha14}, to larger molecules. Using the approach we have calculated REA cross sections for the three negative molecular ions CN$^-$, C$_2$H$^-$, and C$_4$H$^-$. The following concluding remarks should be stressed as a result of the present study.
\begin{itemize}
 \item  
 For completeness of the approach, two pathways for the process have been considered: (1) In the direct  pathway, the electron, incident on the molecule, emits a photon and forms a negative ion in one of the lowest vibrational levels. (2) The indirect  pathway is a two-step process, for which the incident electron is initially captured through  non-Born-Oppenheimer coupling into a vibrationally excited state of the anion, forming a resonance. As a second step in IREA, the resonant vibronic state of the anion emits a photon, which stabilizes the anion with respect to autodetachment. The contribution of the indirect pathway was found to be negligible compared to the direct mechanism if no unusual threshold effects, virtual states or vibrational Feshbach dipole resonances are present.

 \item  
The obtained REA rate coefficients evaluated at temperature 30~K are $7\times 10^{-16}$cm$^3$/s for CN$^-$, $7\times 10^{-17}$cm$^3$/s for  C$_2$H$^-$, $2\times 10^{-16}$cm$^3$/s for C$_4$H$^-$. The coefficients depend weakly on temperature between 10~K and 100~K and increase relatively fast with temperature above 200~K. The validity of the obtained results is verified by comparing the present theoretical results with experimental data from recent photodetachment experiments. 

\item
Previously, it was believed that carbon-chain anions, C$_n$H$^-$ ($n=2,4,6,8$) or C$_m$N$^-$ ($m=3,5$), observed in the interstellar medium, are formed by the REA process. The REA rate coefficients obtained in this study are too small to explain the observed abundance of the anions in the ISM. For example, for C$_4$H$^-$, the magnitude of the rate coefficient needed to explain the observed abundance should be of the order of $10^{-10}$~cm$^3$/s \cite{herbst08}. Thus, the present results suggest that in the ISM, either dipole resonant states or non-local threshold effects increase drastically the REA cross section (and the rate of anion formation) or these anions are formed through a process different than REA. 
\end{itemize}

{\bf Acknowledgments.} This work is supported by the National Science Foundation, Grant No's PHY-11-60611 and  PHY-10-68785. Part of the material presented on this manuscript is based on work conducted while A. Orel was serving at NSF.

\appendix
\section{ Formula for density of states}
\label{app:1}

\subsection{Total number of states}
\label{app:A}

We assume that we have $\cal S$ vibrational modes of the negative ion $M^-$, and the energy of quanta in different modes is approximately the same $\Delta E$. To reach energy $E_i+E_{el}$ of the initial state of the $M+e^-$ system, an excitation of $\cal V$ quanta is needed 
\begin{equation}
\label{eq:sum_partial}
 \sum\limits_{s=1}^{\cal S}v_s = {\cal V}\ \mathrm{with}\ v_s\ge 0\,.
\end{equation}
The density of vibrational states of the $M^-$ anion near energy $E_i+E_{el}$ is evaluated as $N_T/\Delta E$, where $N_T$ is the number of combinations $\{v_1,v_2,\cdots,v_{\cal S}\}$ how the sum above can be formed. 

To obtain the number of combinations $N_T$, it is convenient to represent $\cal V$ quanta as objects arranged in a row. The row of quanta is  separated in subsets  $v_1,v_2,\cdots,v_{\cal S}$ by ${\cal S}-1$ walls.  Now we would like to count the number of possibilities how the ${\cal S}-1$ walls can be placed. We can place the first wall in  ${\cal V}+1$ places between the objects, keeping in mind that we can also place it before the first or after the last quantum. The second wall can be placed at ${\cal V}+2$ different locations, because now we have ${\cal V}$ quanta plus the first wall in the row. (We will account later for the fact that the walls are identical). Continuing in this way, we obtain the total number $\prod\limits_{i=1}^{S-1}({\cal V}+i)=({\cal V+S}-1)!/{\cal V}!$ of possibilities of how the ${\cal S}-1$ distinguishable walls can be placed.  Since the walls are all the same, we have to divide the product by the number $({\cal S}-1)!$ of permutations of the $({\cal S}-1)$ walls for a given partition of the row of ${\cal V}$. Thus, the number of combinations corresponding to the sum of Eq. (\ref{eq:sum_partial}) is  
\begin{equation}
\label{eq:N_T}
N_T={{{\cal V+S}-1}\choose{{\cal S}-1}}={{{\cal V+S}-1}\choose{{\cal V}}}\,.
\end{equation}

\subsection{Number of density of states when only $n$ modes are excited}
\label{app:B}

We derive now the formula for the number of combinations corresponding to $n$ excited modes, given that there are $\cal S$ modes. For example, if only the first $n$ modes are excited, we have 
\begin{equation}
\label{eq:sum_partial_text}
 \sum\limits_{s=1}^{n}v_s = {\cal V}\ \mathrm{with}\ v_s>0\,.
\end{equation}
 In this situation, the number of combinations how the sum above can be formed is given by ${{{\cal V}-1}\choose{n-1}}$. To prove this it is again convenient to represent the $\cal V$ quanta as objects arranged in a row. The row of quanta is  separated in groups of $v_1$, $v_2,\cdots$, $v_n$ objects, by $n-1$ walls.  Now we would like to count the number of possibilities such that $n-1$ walls can be placed. We can place the first wall in  ${\cal V}-1$ places between the objects, keeping in mind that at the left and the right of the wall there were at least one quantum because $v_s>0$. The second wall cannot be placed at same the position, because it would mean that one of the $n$ modes will be inactive, i.e. $v_s=0$. Therefore, for the second wall there are ${\cal V}-2$ ways to place it. The total number of possibilities to place $n-1$ walls is, therefore, $\prod\limits_{i=1}^{n-1}({\cal V}-i)$. Since the walls are all the same, we have to divide the product by the number $(n-1)!$ of permutations of the identical $(n-1)$ walls for a given partition of the row of ${\cal V}$ objects. Thus, the number 
of combinations corresponding to the sum of Eq. (\ref{eq:sum_partial}) is indeed ${{{\cal V}-1}\choose{n-1}}$.

To obtain the total number of combinations $N_n$ when any $n$ modes are excited, we have to multiply the above number with the number of ways how $n$ modes can be chosen from the set of $\cal S$ modes. There are ${{{\cal S}}\choose{n}}$ such ways. Therefore,

\begin{equation}
\label{eq:N_n}
 N_n={{{\cal V}-1}\choose{n-1}}{{{\cal S}}\choose{n}}\,.
\end{equation}

Finally, we will prove that the sum of $N_n$ gives the total number $N_T$ of vibrational states of Eq. (\ref{eq:N_T}), i.e. we will show that
\begin{equation}
\label{eq:N_n_N_T}
 N_T=\sum\limits_{n=1}^{\cal S}N_n\,.
\end{equation}
It can be verified using Vandermonde's identity, which states that
\begin{equation}
\label{eq:vandermonde}
{{m+l}\choose{r}}=\sum\limits_{n=0}^r{{m}\choose{n}}{{l}\choose{r-n}}\,.
\end{equation}
In our case, we have
\begin{eqnarray}
\label{eq:N_n_N_T2}
\sum\limits_{n=1}^{\cal S}N_n&=& \sum\limits_{n=1}^{\cal S} {{{\cal V}-1}\choose{n-1}}{{{\cal S}}\choose{n}} = \sum\limits_{n=0}^{\cal S'} {{{\cal V}-1}\choose{n}}{{{{\cal S'}+1}}\choose{n+1}} \nonumber\\
&=& \sum\limits_{n=0}^{\cal S'} {{{\cal V}-1}\choose{n}}{{{{\cal S'}+1}}\choose{{\cal S'}-n}} \,,
\end{eqnarray}
where ${\cal S'}={\cal S}-1$. Now using Vandermonde's identity with $m={\cal V}-1$, $l={\cal S'}+1$ and $r=\cal S'$, the sum is written
\begin{equation}
\label{eq:N_n_N_T_final}
\sum\limits_{n=1}^{\cal S}N_n ={{{\cal V+S}-1}\choose{{\cal S}-1}} \,,
\end{equation}
which is indeed $N_T$ of Eq. (\ref{eq:N_T}).

\section{Closed-form expression for ${\mathcal I}({\cal V}_0)$ and statistical formula}
\label{app:2}
We will reduce ${\mathcal I}({\cal V}_0)$ in Eq. (\ref{eq:sum}) to a closed-form expression. In order to simplify the equations, we introduce the shorthand notation ${\bar{\mathcal I}}_0=2{\mathcal I}({\cal V}_0)\exp({\beta_0^2})$. From Eq. (\ref{eq:lambda}), we decompose ${\bar{\mathcal I}}_0$ in three terms:
\begin{eqnarray}
{\bar{\mathcal I}}_0&=&\sum_{s=1}^{\cal S}\sum_{ {\cal V}_f={\cal V}_0}\frac{\beta_1^{2v_1}\cdots\beta_s^{2v_s}\cdots\beta_{{\cal S}}^{2v_{{\cal S}}}}{v_1!\cdots v_s!\cdots v_{{\cal S}}!}\nonumber\\
&&\times(\underbrace{\beta_s^2}_{1}-\underbrace{2v_s}_{2}+\underbrace{v_s^2/\beta_s^2}_{3})\,.
\end{eqnarray}
The under braces indicate the factors involved in each term, such that ${\bar{\mathcal I}}_0={\bar{\mathcal I}}_1+{\bar{\mathcal I}}_2+{\bar{\mathcal I}}_3$. For convenience, we introduce the function
\begin{eqnarray}
{\bar{\mathcal Z}(\beta_1\cdots\beta_{{\cal S}})}\equiv\sum_{ {\cal V}_f={\cal V}_0}\frac{\beta_1^{2v_1}\cdots\beta_s^{2v_s}\cdots\beta_{{\cal S}}^{2v_{{\cal S}}}}{v_1!\cdots v_s!\cdots v_{{\cal S}}!}\,.
\end{eqnarray}
Recalling the multinomial theorem, this function takes the compact form
\begin{eqnarray}
\label{eq:F}
{\bar{\mathcal Z}(\beta_1\cdots\beta_{{\cal S}})}=\frac{(\beta_1^2+\cdots+\beta^2_{\cal S})^{{\cal V}_0}}{{\cal V}_0!}=\frac{\beta_0^{2{\cal V}_0}}{{\cal V}_0!}\,.
\end{eqnarray}
The first term ${\bar{\mathcal I}}_1$ is expressed as
\begin{eqnarray}
{\bar{\mathcal I}}_1=\sum_{s=1}^{\cal S}\beta_s^2{\bar{\mathcal Z}(\beta_1\cdots\beta_{{\cal S}})}\,,
\end{eqnarray}
such that we readily obtain
 \begin{eqnarray}
{\bar{\mathcal I}}_1=\frac{\beta_0^{2{\cal V}_0+2}}{{\cal V}_0!}\,.
\end{eqnarray}
The second term ${\bar{\mathcal I}}_2$ is written as
\begin{eqnarray}
{\bar{\mathcal I}}_2=-\sum_{s=1}^{\cal S}\sum_{ {\cal V}_f={\cal V}_0}2v_s\frac{\beta_1^{2v_1}\cdots\beta_s^{2v_s}\cdots\beta_{{\cal S}}^{2v_{{\cal S}}}}{v_1!\cdots v_s!\cdots v_{{\cal S}}!}\,,
\end{eqnarray}
which takes the convenient form
\begin{eqnarray}
{\bar{\mathcal I}}_2=-\sum_{s=1}^{\cal S}\frac{\partial \bar{\mathcal Z}(\beta_1\cdots\beta_{{\cal S}})}{\partial\ln\beta_s}\,.
\end{eqnarray}
Using the value of ${\bar{\mathcal Z}(\beta_1\cdots\beta_{{\cal S}})}$  in Eq. (\ref{eq:F}), we obtain the following expression:
\begin{eqnarray}
{\bar{\mathcal I}}_2=-\frac{2{\cal V}_0\beta_0^{2{\cal V}_0}}{{\cal V}_0!}\,.
\end{eqnarray}
The last term ${\bar{\mathcal I}}_3$ is clearly the dominant one, with the slightly more complicated form:
\begin{eqnarray}
\label{eq:I3}
{\bar{\mathcal I}}_3=\sum_{s=1}^{\cal S}\sum_{ {\cal V}_f={\cal V}_0}\frac{v^2_s}{\beta^2_{s}}\frac{\beta_1^{2v_1}\cdots\beta_s^{2v_s}\cdots\beta_{{\cal S}}^{2v_{{\cal S}}}}{v_1!\cdots v_s!\cdots v_{{\cal S}}!}\,.
\end{eqnarray}
Note that the apparent indeterminacy is easily removed for a degenerate mode by taking the limit $\beta_s\to0$ in Eq. (\ref{eq:I3}). Applying a similar method, we find
\begin{eqnarray}
{\bar{\mathcal I}}_3=\frac{1}{4}\sum_{s=1}^{\cal S}\frac{1}{\beta^2_{s}}\frac{\partial^2{\bar{\mathcal Z}(\beta_1\cdots\beta_{{\cal S}})}}{\partial (\ln \beta_s)^2}\,,
\end{eqnarray}
and thus we conclude that
\begin{eqnarray}
{\bar{\mathcal I}}_3=\frac{{\cal V}_0({\cal V}_0+{\cal S}-1)\beta_0^{2{\cal V}_0-2}}{{\cal V}_0!}\,.
\end{eqnarray}
Finally, the closed-form formula for ${\mathcal I}({\cal V}_0)$ is given by:
\begin{eqnarray}
{\mathcal I}({\cal V}_0)=\frac{\beta_0^{2{\cal V}_0-2}\left[({\cal V}_0-\beta_0^2)^2+{\cal V}_0({\cal S}-1)\right]}{{\cal V}_0!}\frac{e^{-\beta_0^2}}{2}.
\end{eqnarray}
In the above expression, $\beta^2_0\ll{\cal V}_0$, which allows us to simplify the expression as 
\begin{eqnarray}
\label{eq:I_factor_appendix}
{\mathcal I}({\cal V}_0)\approx\frac{({\cal V}_0+{\cal S}-1)}{({\cal V}_0-1)!}\frac{\beta_0^{2{\cal V}_0-2} e^{-\beta_0^2}}{2}\,.
\end{eqnarray}

We also derive an expression for the average total energy of the system, weighted over the electron capture transition probabilities, given that ${\cal V}_0$ quanta are excited. First, neglecting the terms $\beta_s^2\ll v_s$ in Eq. (\ref{eq:lambda}), the transition probabilities to capture in an anion vibrational state are  proportional to  
\begin{eqnarray}
{\cal P}(\beta_1\cdots\beta_{\cal S})=v_s\frac{\beta_1^{2v_1}\cdots\beta_s^{2v_s-2}\cdots\beta_{{\cal S} }^{2v_{{\cal S} }}}{v_1!\cdots (v_s-1)!\cdots v_{{\cal S} }!}.
\end{eqnarray}
Therefore, the partition function used to normalize the probabilities takes the form
\begin{eqnarray}
 \Xi=\sum_{s=1}^{\cal S}\sum_{ {\cal V}_f={\cal V}_0}{\cal P}(\beta_1\cdots\beta_{\cal S})=\bar{{\cal I}}_3\,.
\end{eqnarray}
and the average energy is thus expressed as
\begin{eqnarray}
\langle E\rangle&=&\frac{1}{\Xi}\sum_{{\cal V}_f={\cal V}_0}E_f{\cal P}(\beta_1\cdots\beta_{\cal S})\,,
\end{eqnarray}
where $E_f=\hbar(v_1\omega_1+\cdots+v_{{\cal S} }\omega_{{\cal S} })$ is the energy of the vibrational state $|\chi_f\rangle$ (the anion ground state is chosen as the reference energy). After some straightforward manipulations, we obtain for ${\cal V}_0>1$ the expression
\begin{eqnarray}
\label{eq:av_energy_final}
\langle E\rangle&\approx&({\cal V}_0-1)\hbar\langle\omega\rangle,
\end{eqnarray}
where $\langle\omega\rangle=\frac{1}{\beta_0^2}\sum\beta_s^2\omega_s$ is an average vibrational frequency weighted over the displacements $\beta_s$. In a similar fashion, one can calculate the energy spread $\Delta E=\sqrt{\langle E^2\rangle-\langle E\rangle^2}$ around the average energy $\langle E\rangle$, which takes the value 
\begin{eqnarray}
\Delta E\approx\hbar\Delta\omega\sqrt{{\cal V}_0-1},
\end{eqnarray}
where $\Delta\omega=\sqrt{\langle\omega^2\rangle-\langle\omega\rangle^2}$ is the vibrational frequency spread, with $\langle\omega^2\rangle=\frac{1}{\beta_0^2}\sum\beta_s^2\omega^2_s$.


\end{document}